\documentclass[twocolumn]{aastex631}

\received{December 20, 2022}
\revised{June 30, 2023}
\accepted{July 11, 2023}

\submitjournal{AJ}

\shorttitle{SALT Fabry-P\'erot Observations of NGC\,1068}
\shortauthors{Hviding et al.}

\graphicspath{{figures}}

\usepackage{amsmath}

\begin{document}

\title{The Kiloparsec Scale Influence of the AGN in NGC\,1068 with SALT RSS Fabry-P\'erot Spectroscopy\footnote{{based on observations made with the Southern African Large Telescope (SALT)}}}

\correspondingauthor{Raphael E. Hviding}
\email{rehviding@email.arizona.edu}

\author[0000-0002-4684-9005]{Raphael E. Hviding}
\affiliation{Steward Observatory, University of Arizona, 933 North Cherry Avenue, Rm. N204, Tucson, AZ 85721, USA}

\author[0000-0003-1468-9526]{Ryan C. Hickox}
\affiliation{Department of Physics and Astronomy, Dartmouth College, 6127 Wilder Laboratory, Hanover, NH 03755, USA}

\author[0000-0001-7673-4850]{Petri V\"{a}is\"{a}nen}
\affiliation{South African Astronomical Observatory, P.O. Box 9 Observatory, Cape Town, South Africa}

\author{Rajin Ramphul}
\affiliation{South African Astronomical Observatory, P.O. Box 9 Observatory, Cape Town, South Africa}

\author[0000-0003-4565-8239]{Kevin N. Hainline}
\affiliation{Steward Observatory, University of Arizona, 933 North Cherry Avenue, Rm. N204, Tucson, AZ 85721, USA}
\begin{abstract}

    We present Fabry-P\'erot (FP) imaging and longslit spectroscopy of the nearby Seyfert II galaxy NGC\,1068 using the Robert Stobie Spectrograph (RSS) on the Southern African Large Telescope (SALT) to observe the impact of the central Active Galactic Nucleus (AGN) on the ionized gas in the galaxy on kiloparsec scales. With SALT RSS FP we are able to observe the H$\alpha$+[\ion{N}{2}] emission line complex over a $\sim$2.6\,arcmin$^2$ field of view. Combined with the longslit observation, we demonstrate the efficacy of FP spectroscopy for studying nearby Type II Seyfert galaxies and investigate the kiloparsec-scale ionized gas in NGC\,1068. We confirm the results of previous work from the TYPHOON/Progressive Integral Step Method (PrISM) survey that the kiloparsec-scale ionized features in NGC\,1068 are driven by AGN photoionization. We analyze the spatial variation of the AGN intensity to put forward an explanation for the shape and structure of the kiloparsec-scale ionization features. Using a toy model, we suggest the ionization features may be understood as a light-echo from a burst of enhanced AGN activity $\sim$2000 years in the past. 

\end{abstract}


\section{Introduction} \label{sec:intro}

Since the discovery of quasars, the most luminous Active Galactic Nuclei (AGNs), in the early 1960s \citep{schmidt3C273StarLike1963}, a multitude of AGN classifications have arisen over the last few decades to form the ``AGN Zoo'' \citep[e.g.][]{padovaniActiveGalacticNuclei2017}. AGN unification attempts to explain the different classifications through the geometry of the obscuring material surrounding the central supermassive black hole (SMBH) and accretion disk \citep[e.g.][]{antonucciUnifiedModelsActive1993,urryUnifiedSchemesRadioLoud1995,netzerRevisitingUnifiedModel2015}.  AGNs can be divided into two populations: obscured sources with narrow high-ionization nebular emission lines (Type II), and unobscured sources with additional broad (FWHM\,$>$1000\,km\,s$^{-1}$) emission lines (Type I). In the standard unification model, these classifications correspond to different amounts of dust along our line of sight, due to the orientation of a parsec-scale obscuring torus, where sightlines that avoid the obscuring material can observe the fast moving gas closer to the central nucleus \citep[e.g.][]{netzerRevisitingUnifiedModel2015,ramosalmeidaNuclearObscurationActive2017}.

However, recent work suggests that Type II AGN, classically understood as a difference in line of sight to the central engine, have been found to live in different environments and generate more powerful outflows \citep[e.g.,][]{dipompeoAngularClusteringInfraredselected2014,dipompeoUpdatedMeasurementsDark2016,dipompeoIIIProfilesInfraredselected2018,zakamskaDiscoveryExtremeIII2016,zakamskaHostGalaxiesHighredshift2019,mitraHaloOccupationDistribution2018}, suggesting they differ from their unobscured counterparts by more than just viewing angle. An evolutionary paradigm has emerged, proposing that dynamical processes in gas-rich galaxies drive obscuring material into the nuclear region, producing the observed optical and ultraviolet (UV) attenuation \citep[e.g.][]{kauffmannUnifiedModelEvolution2000,hopkinsUnifiedMergerdrivenModel2006,hopkinsDissipationExtraLight2008}. In this case the merger also drives gas into the nucleus to fuel AGN, producing the feedback that eventually remove the dust and gas and potentially affecting the galaxy on larger scales \citep{fabianObservationalEvidenceActive2012,alexanderWhatDrivesGrowth2012}. The study of obscured AGN can shed light on the fueling processes and the galactic scale effects of AGN feedback \citep[for a review, see][and references therein]{hickoxObscuredActiveGalactic2018}.  

Moderate luminosity Type II AGNs (i.e. Seyfert II galaxies) are ideal to examine the large-scale effects of an obscured AGN on its host. The lack of broad optical and UV emission lines imply obscuration of high velocity gas in the nuclear region. In many previous studies, spectroscopic analyses of Type II AGNs have used line diagnostic techniques to understand the nature of AGN emission integrated over galaxy scales. Recently, integral field units (IFUs) have enabled spatially-resolved spectral diagnostics for a large number of galaxies, but are generally limited to small fields of view \citep[e.g.][]{baconMUSESecondgenerationVLT2010,croomSydneyAAOMultiobjectIntegral2012,bundyOverviewSDSSIVMaNGA2015}. Therefore it has been challenging to obtain spatially-resolved spectral diagnostics of the best-studied nearby AGN that extend over large angular scales. 

A powerful tool for probing the spectra of galaxies over a large field of view (FoV) is Fabry-P\'erot (FP) spectroscopy, which samples the source with extremely narrow bandpasses generated by precisely spacing an etalon for each exposure. {FP spectroscopy can therefore provide moderate resolution spectroscopy across large FoVs without sacrificing spatial resolution. The technique is especially powerful when paired with large aperture telescopes which can obtain the necessary sensitivity for detailed studies of nearby galaxies with large angular extents.}

By measuring the spectroscopic properties of galaxies over a large physical extent,  it is possible to reveal the presence of extended emission-line regions on $\sim$kpc scales. These emission line regions can be indicative the current accretion luminosity from the black hole \citep{kauffmannHostGalaxiesActive2003,lamassaSDSSIVEBOSSSpectroscopy2019} but may be due to {\em past} enhanced AGN activity that can manifest as light or ionization ``echoes'' which have been studied extensively in \citet{keelGalaxyZooSurvey2012,keelHSTImagingFading2015,keelFadingAGNCandidates2017}. These echoes can be used to constrain the accretion history of the AGN and potentially shed light on the coevolution of the central SMBH and its host galaxy.

In this work, we obtain FP spectroscopy from the Robert Stobie Spectrograph \citep[RSS;][]{burghPrimeFocusImaging2003,nordsieckInstrumentationHighresolutionSpectropolarimetry2003,kobulnickyPrimeFocusImaging2003,smithPrimeFocusImaging2006} with the Southern African Large Telescope (SALT) of the nearby Seyfert II galaxy NGC\,1068 in order to observe the impact of AGN on ionized gas in the galaxy on kiloparsec scales.  SALT RSS FP spectroscopy affords a large FoV along with the necessary spatial and spectral resolution to obtain high-quality velocity and ionization maps for the line-emitting gas across the extent of this galaxy.

We describe NGC\,1068 and our observations in Section \ref{sec:ngc} and in Section \ref{sec:reduction} we detail the reduction of our data. We outline the procedure for producing line emission diagnostic maps of NGC\,1068 from our reduced data in Section \ref{sec:analysis}. We present the final diagnostic maps and derived results in Section \ref{sec:results} along with our interpretation of the results. Finally, our we discuss our results in Section \ref{sec:conclude} and discuss potential future research.

\section{NGC\,1068 and Observations}
\label{sec:ngc}

NGC\,1068, also known as Messier\,77, is a barred spiral galaxy and the optically brightest southern Seyfert galaxy known \citep{devaucouleursSouthernGalaxiesVI1973}. As a Type II AGN, NGC 1068 presents possibly the best opportunity to conduct resolved studies of the interaction between a galaxy and obscured SMBH growth. 
Indeed, the obscuring material around the galaxy's central engine has been the subject of intense study for the past few decades. The nuclear region has been examined in the radio with Very Long Baseline Interferometry from the Very Large Array and in the infrared using interferometry from the Very Large Telescope to glean insight into the distribution, composition, and thermal properties of the obscuring material \citep{vanderhulstVLAObservationsSeyfert1982,greenhillVLBIImagingWater1996,rottgeringObservingSeyfertNucleus2004,gamezrosasThermalImagingDust2022}. 

NGC\,1068 has also been studied with the Atacama Large Millimeter Array (ALMA), which has been able to resolve the dusty torus around the SMBH on relatively small (parsec) scales \citep{garcia-burilloALMAResolvesTorus2016}. The Hubble Space Telescope (HST) obtained high-resolution spectral information using the Space Telescope Imaging Spectrograph (STIS) in order to constrain the dynamics of the inner 400~pc of the system \citep{dasKinematicsNarrowLineRegion2006,dasDynamicsNarrowLineRegion2007}. Longslit studies in the infrared have covered similar regions to measure the photoionization properties of the galaxy nucleus \citep[e.g.][]{tamuraNearinfraredLongslitSpectra1991,martinsNuclearExtendedSpectra2010}.

{Wider-field observations, narrow-band imaging and spectroscopy have demonstrated the existence of extended emission on kiloparsec scales from the central nucleus, with heightened [\ion{O}{3}], [\ion{N}{2}], and Balmer line emission seen in \citet{baldwinKinematicsIonizationExtended1987} and \citet{poggeExtendedIonizingRadiation1988}.} The innermost portion of these regions is also covered by Multi Unit Spectroscopic Explorer (MUSE) optical IFU observations from the Measuring Active Galactic Nuclei Under MUSE Microscope (MAGNUM) survey, where the ionizing source behind the emission is consistent with AGN activity  \citep{venturiMAGNUMSurveyCompact2021}. The TYPHOON/Progressive Integral Step Method (PrISM) survey has extended the range of observations with a stepped longslit program using the Wide Field reimaging CCD imaging spectrograph on the du Pont telescope at the Las Campanas Observatory, {confirming the regions as AGN likely bring driven by AGN ionization several kiloparsecs} from the galaxy nucleus \citep{dagostinoStarburstAGNMixingTYPHOON2018}. 

\begin{figure*}[ht!]
    \centering
    \gridline{
        \fig{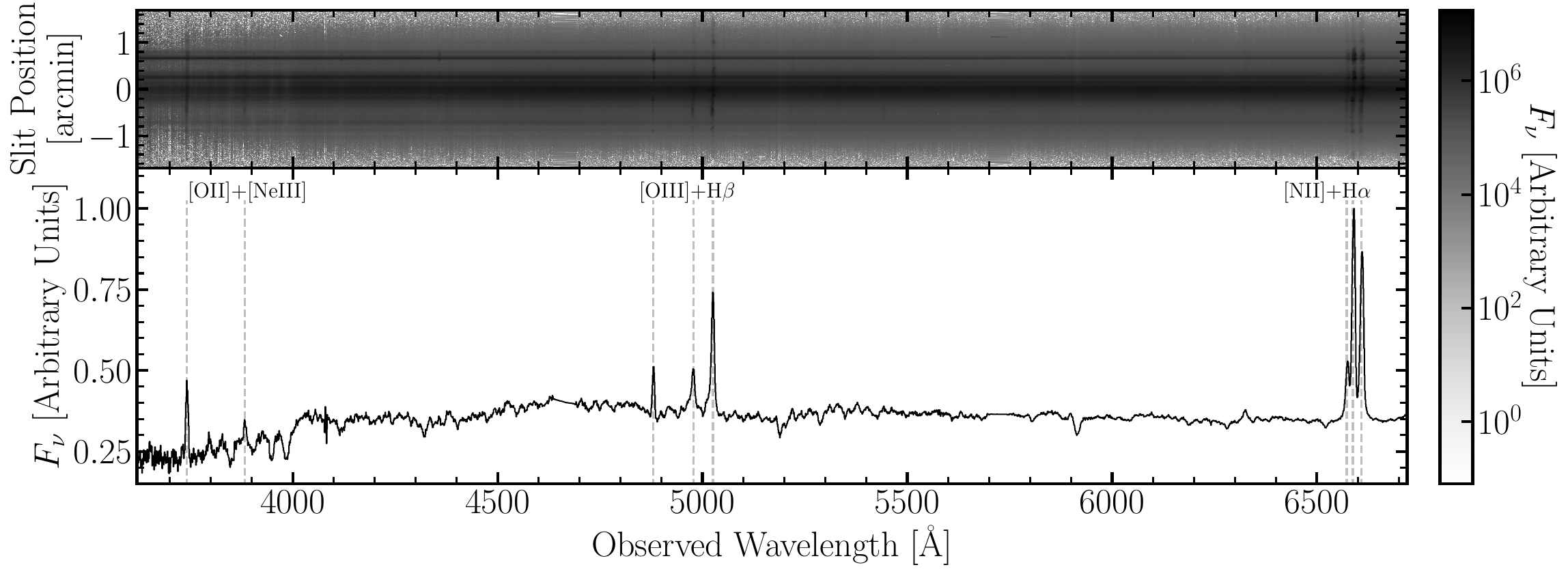}{\textwidth}{(a) Longslit}
    }
    \gridline{
        \fig{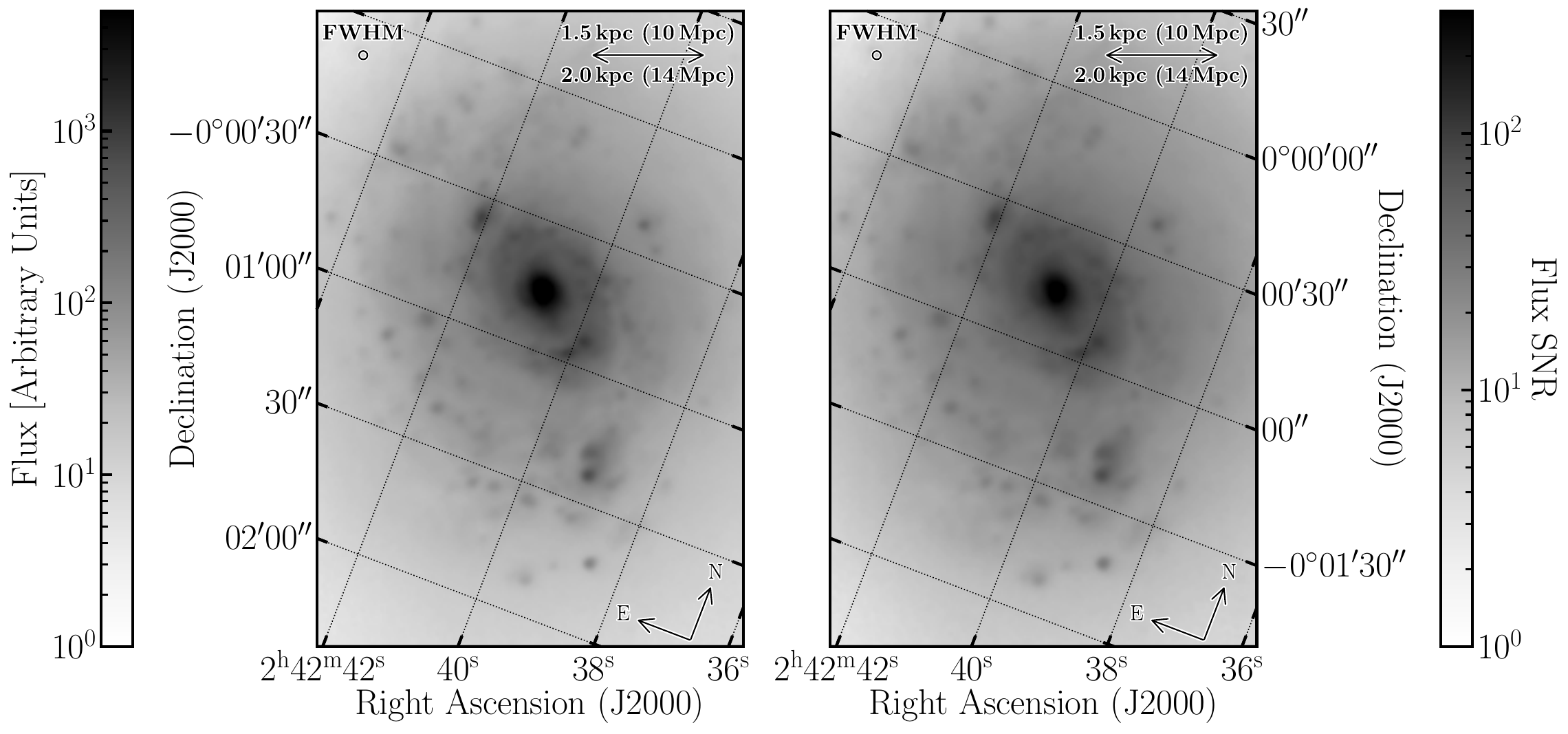}{\textwidth}{(b) Fabry-P\'erot}
    }
    \caption{The reduced data of NGC\,1068. In panel (a) we present the reduced 2D spectrum of NGC\,1068 taken with RSS longslit (top) along with the sum along the spatial dimension (bottom). In panel (b) we show the total measured intensity of the FP observations {(right) along with the associated SNR (left)}. {In addition we provide the spatial scale of the image assuming 10.1\,Mpc and 13.97\,Mpc distances to NGC\,1068 (top right), the orientation of the image (bottom right), and a circle depicting the final FWHM of our observation.} Due to the nature of the stationary primary mirror of SALT, no absolute flux calibration can be performed from the data.\label{fig:intensity}}
\end{figure*}

We obtained RSS FP spectroscopy with SALT on 2015-11-09 under proposal ID 2015-2-SCI-024. FP spectroscopy is ideal to study the kiloparsec scale ionized regions in NGC\,1068. We make use of the FP spectroscopy mode with RSS on SALT described in \citet{rangwalaImagingFABRYPEROTSystem2008}. We utilized the Low Resolution etalon, which affords a $\mathcal{R}$ of 779, a free spectral range of 182\,\AA, and a finesse of 21.9, all quoted at 6500\AA\ {with an 8\arcmin\ diameter FoV}.
{Our target wavelength range is selected to target the H$\alpha$+[\ion{N}{2}]$\lambda\lambda$6583,6548\AA\ complex. Adopting a systemic redshift of $z = 0.003793$ \citep[1137\,km\,s$^{-1}$;][]{huchraCFARedshiftSurvey1999} for NGC\,1068 we therefore target the observed wavelength range of 6570\AA\ to 6620\AA\ corresponding to a rest wavelength range of 6545\AA\ to 6595\AA}.

{We perform 2 integrations at 13 etalon positions with a PA of 159$^\circ$ and an exposure time of 20 seconds. Each etalon position is set such that the transmitted wavelength is 4\AA\ greater than the previous setting, so that we are appropriately sampling our $\sim$8\AA\ spectral resolution}. Since the exact transmitted bandpass is a function of the position in the FoV, the spectral coverage of each region of the image varies on the order of a few \AA. In order to account for the chip gaps, a single dither is performed and all etalon spacings are repeated for a total of {1040\,sec of science exposure time comprised of 52 frames}.

In addition, we make use of a previously observed RSS longslit spectrum of NGC\,1068 to validate our reduction and analysis of our FP data. The longslit observation was taken on 2011-07-30 under proposal ID 2011-2-RSA\_OTH-002 using the PG0900 grating with a grating angle of 13.62$^\circ$ {affording an $\mathcal{R}$ of $\sim1400$ and $\sim1000$ at 6600\AA\ and 5000\AA\ respectively covering the H$\alpha$+[\ion{N}{2}] and H$\beta$+[\ion{O}{3}]$\lambda\lambda$5007,4959\AA\ complexes}. {A slit 8\arcmin\ long, 2\arcsec\ wide, and a PA of 212$^\circ$ were used for three 300\,sec exposures for a total integration time of 900\,sec.}

{As the distance to NGC\,1068 has remained a subject of debate for decades, in this work we consider the following three distance estimates: (1) 10.1\,Mpc (2.9\,kpc\,arcmin$^{-1}$) using the Tully-Fisher (TF) relation from  \citet{tullyExtragalacticDistanceDatabase2009}, (2) 11.1\,Mpc (3.2\,kpc\,arcmin$^{-1}$) measured from the tip of the red giant branch (TRGB) in \citet{tikhonovTRGBDistancesSeyfert2021}, and (3) 13.97\,Mpc (4.1\,kpc\,arcmin$^{-1}$) derived from numerical action methods (NAM) in \citet{anandDistancesPHANGSGalaxies2021}. We adopt the TF distance method due to it's prevalence in the literature but our analysis will also consider the NAM distance as it is the most recent estimate of the galaxy's distance and the furthest distance we consider.}

\section{Data Reduction}
\label{sec:reduction}

In this section we detail the reduction of our SALT RSS data in both the FP and longslit modes. {As SALT was built using a fixed primary mirror design, for which the effective pupil size changes during the observation, it is difficult to determine absolute flux incident onto the CCD. A relative flux is therefore produced at the end of RSS data reduction for both our longslit and FP data. While it would be possible to perform absolute flux calibration by comparing to existing data we do not do so in this work as our analysis primarily involves measuring the ionization state of the gas which relies on relative flux ratios}.

\subsection{RSS Longslit Spectroscopy}

The longslit spectrum {is reduced using the PySALT\footnote{http://pysalt.salt.ac.za/} user package which serves as the primary reduction software for data from SALT \citep{crawfordPySALTSALTScience2010}}. The frames are bias subtracted, flat-fielded, and a wavelength solution was generated from arc lamp spectra and applied to the data. {The final flux is taken as the median of the frames while the error is computed as the standard deviation between the frames. Finally we apply an interstellar extinction correction to the 2D spectrum a long with a flux calibration generated from spectroscopic standards}. The reduced 2D spectrum and total summed intensity spectrum across the FoV are presented in Figure \ref{fig:intensity}(a).

\subsection{RSS Fabry-P\'erot Spectroscopy} \label{subsec:fpred}

The FP images are reduced following the \texttt{saltfppipe}\footnote{http://saltfppipe.readthedocs.io/en/latest/index.html} pipeline. {The images are bias corrected, flat-fielded, and are mosaiced using the PySALT user package. Corresponding variance images are created based on the pixel variance and the readnoise. Due to the additional complexity of reducing FP data, we describe the process in depth in this section.}

Since each step along the etalon comprises a separate observation, the observing conditions are subject to change over the course of the total acquisition. In order to combine the observations, the average seeing must be measured. Each image can then be convolved with the appropriate function, ensuring that the effective beam size is constant across the observations. The final beam size will be equivalent to the worse seeing across all the images. The effective seeing for each observation is measured by manually selecting several stars across the field, fitting a Gaussian profile to each of them across all images, and comparing their observed FWHM. {The final FWHM for our FP observations of NGC\,1068 was 2\arcsec\ which is shown on all Figures presenting FP data}.

Again, due to the nature of the fixed SALT primary mirror, the illumination fraction onto the primary will change between observations. For this reason, a number of stars are algorithmically detected throughout the image and their relative intensities and positions are recorded. By averaging over the measured intensities, each image can be normalized accordingly. The measured positions also allows for any systematic drift between the images to be corrected through realignment.  
In order to properly generate the wavelength calibration of the observations, it is imperative to discern the optical center of the image. As the SALT RSS FP system induces a reflection, the optical center can be deduced by finding associated reflection pairs. This has the added benefit of being able to mitigate the effects of these ``ghost'' reflections by attempting to account for the effects of the reflection or, failing that, masking the brightest ghost features.

The wavelength calibration changes as a function of the distance from the optical center, the etalon spacing, and time. Given the wavelength stability of RSS FP spectroscopy, {which incurs a wavelength drift $\sim1$\AA\,hour$^{-1}$ and validated to not exceed  FWHM/3\,hour$^{-1}$ in commissioning, \citep{rangwalaImagingFABRYPEROTSystem2008,buckleyCommissioningSouthernAfrican2008}}, we do not introduce any time dependence in our wavelength solution.

Traditional arc lamp spectra allow us to find a precise wavelength at a given radius from the optical center. For SALT observations, two arc lamp spectra are taken at the start and the end of the observation, at the maximum and minimum of the etalon spacing. In addition, the individual images are used in solving the wavelength solution. By subtracting the median image from each observation, the night sky emission rings become prominent and can be matched to known emission line values in the range of the interference filter. 

Following the wavelength calibration, the azimuthally-averaged radial profile for each image is examined and fitted in order to remove the contribution from night sky emission lines. This must be done for each image individually. Finally, the data cube can be assembled from the constituent images, here each observation is convolved with the appropriate function in order to produce the determined beam size. Finally, the pipeline corrects for the telescope's heliocentric velocity after the creation of the cube. {We note that for one frame we were not able to achieve successful subtraction of the sky emission feature and therefore exclude it from the subsequent analysis}.

In order to combine different dithers, a standard approach cannot be taken. Simply averaging pairs of corresponding pixels from the two corresponding images with the same etalon spacings is not advisable with RSS FP data, as two images from differing data cubes will have the object of interest at a different position relative to the optical center, changing the wavelength solution at each position. After discerning the offset, a new data cube is created by combining all of the dither information. Each spectral pixel (spaxel) will simply have twice the number of wavelength-intensity pairs as before, while those spaxels that lie in one of the chip gaps from one dither will have data from the other dither. Finally, stars in the in the FoV are identified in order to generate a World Coordinate System (WCS) transformation between pixel coordinates and the J2000 reference frame. 

The total summed intensity of the entire data cube is presented in Figure \ref{fig:intensity} (b) along with the associated SNR in Figure \ref{fig:intensity} (c). {Emission line regions are highlighted due to the relatively narrow bandpass across our FP observations. In addition, there do not appear to be any spatial variations in noise apart from those induced by fluctuations in the galaxy's brightness.}

\section{Data Analysis}
\label{sec:analysis}

In this section we discuss the analysis techniques performed on the reduced RSS FP spectroscopy data cubes and RSS longslit 2D spectrum in order to extract the emission line fluxes and velocities over the FoV.

\begin{figure*}[ht!]
    \centering
    \gridline{
        \fig{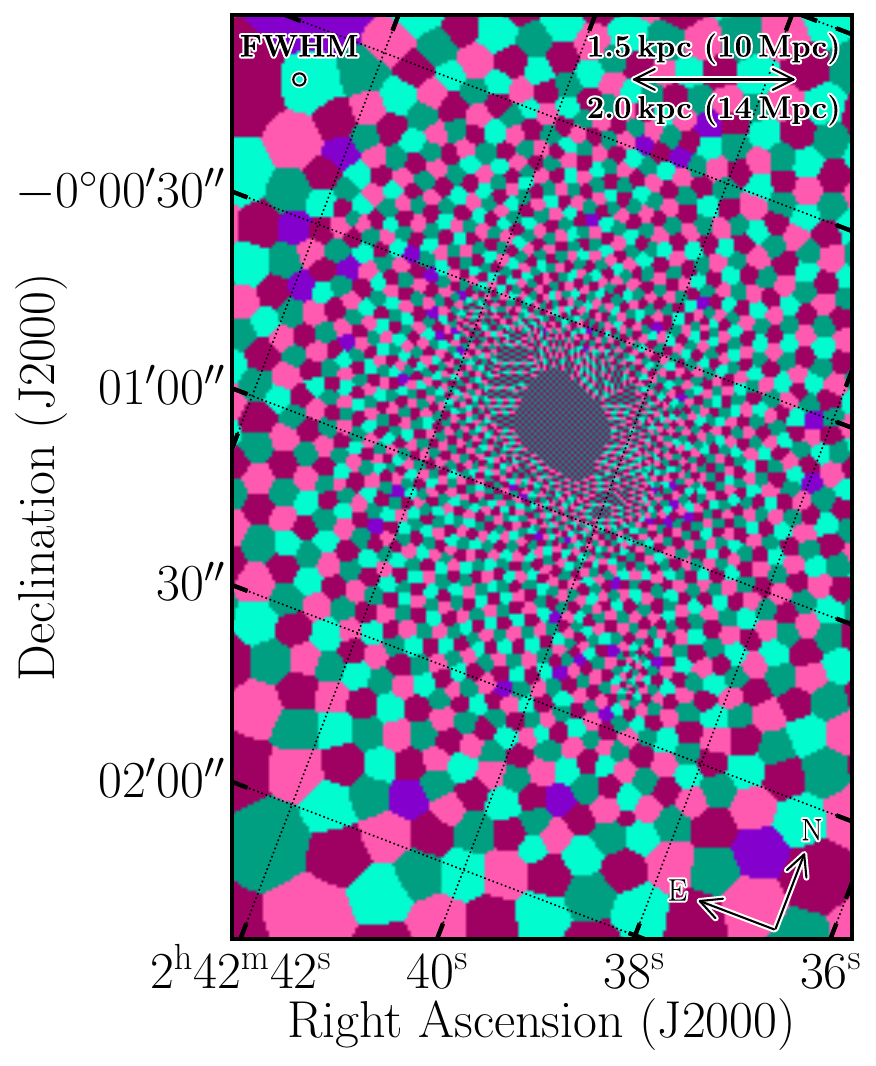}{0.28\textwidth}{(a) Voronoi Tessalation}
        \fig{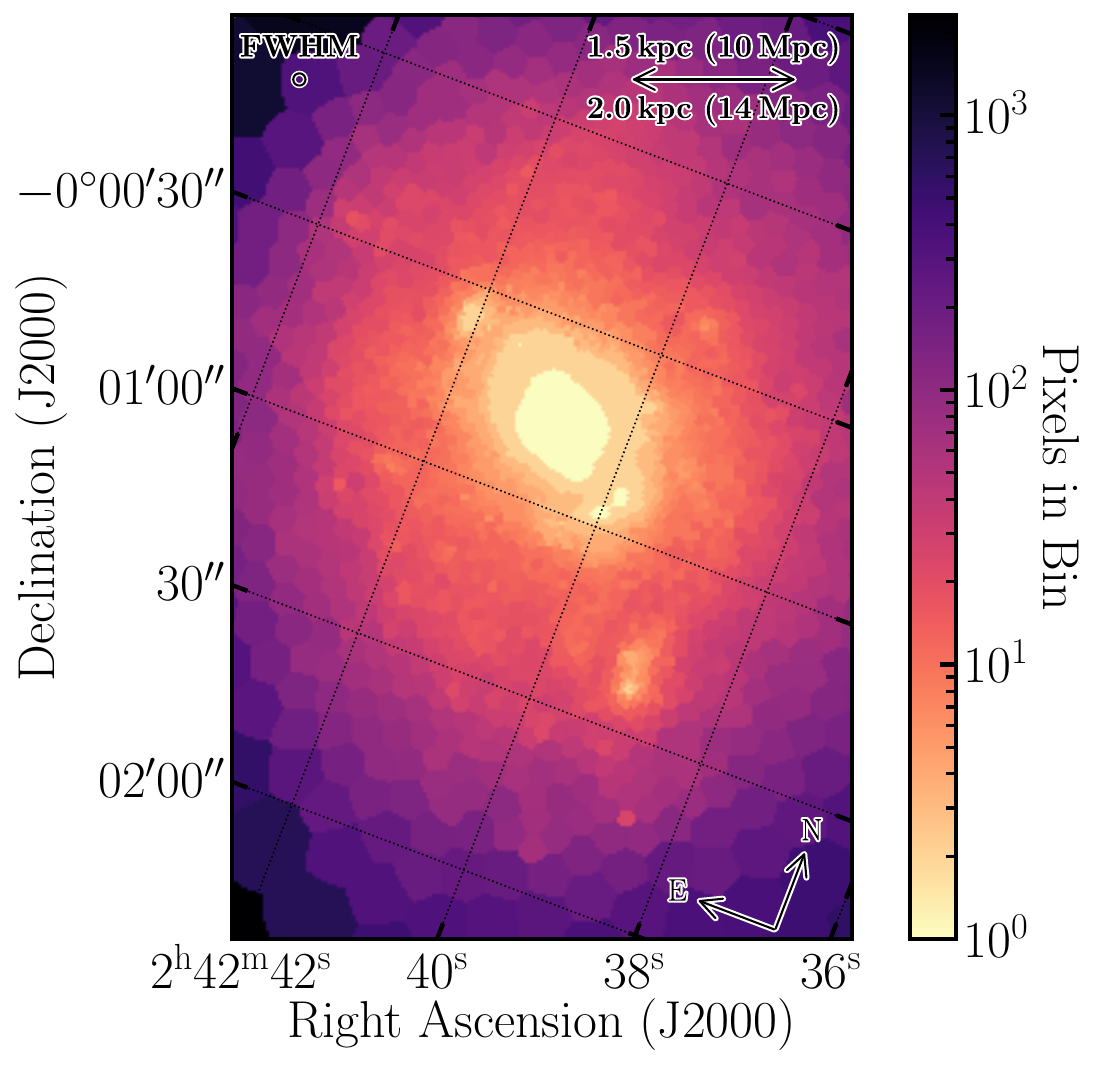}{0.35\textwidth}{(b) Voronoi Number}
        \fig{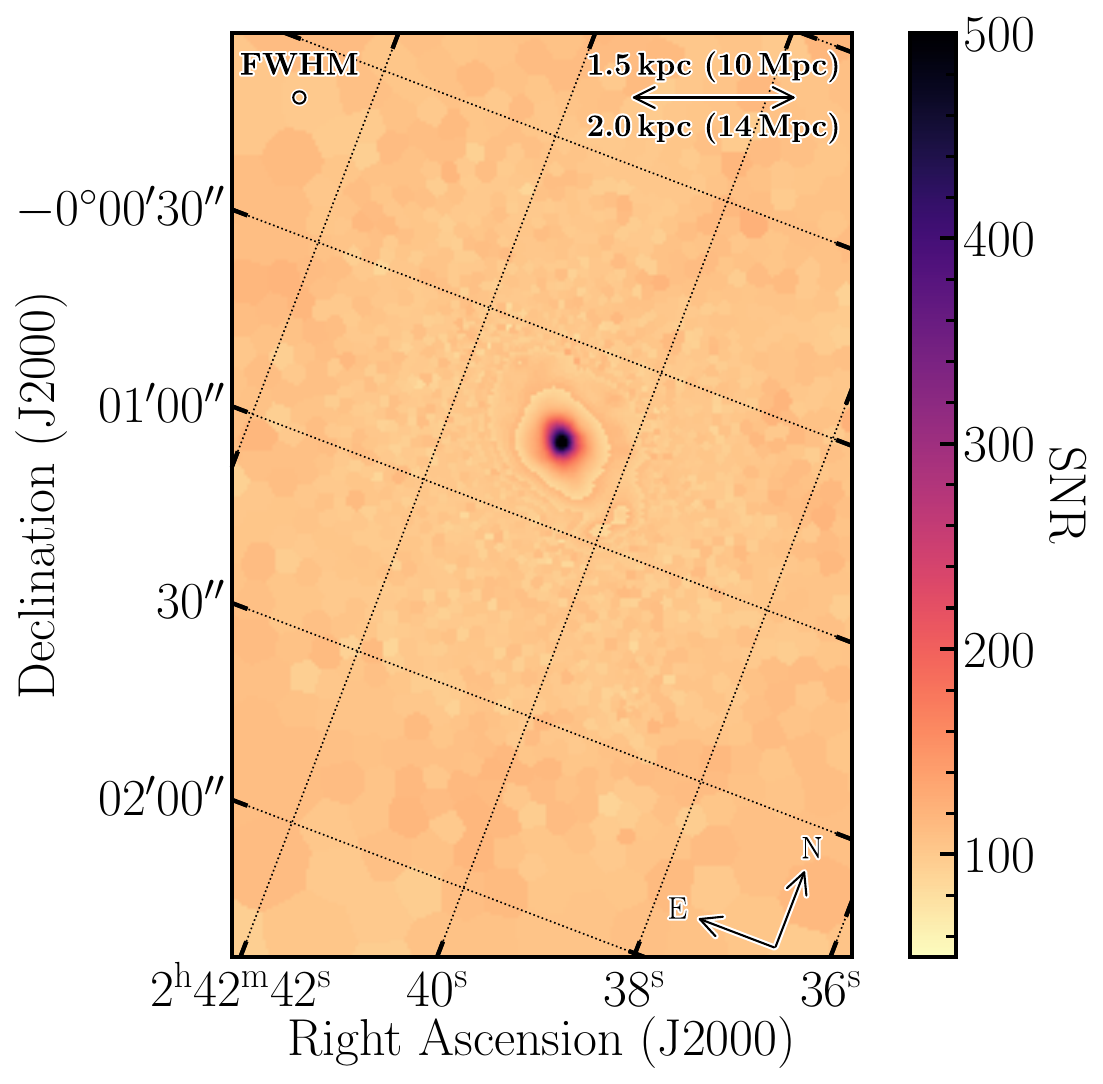}{0.35\textwidth}{(c) Voronoi SNR}
    }
    \gridline{
        \fig{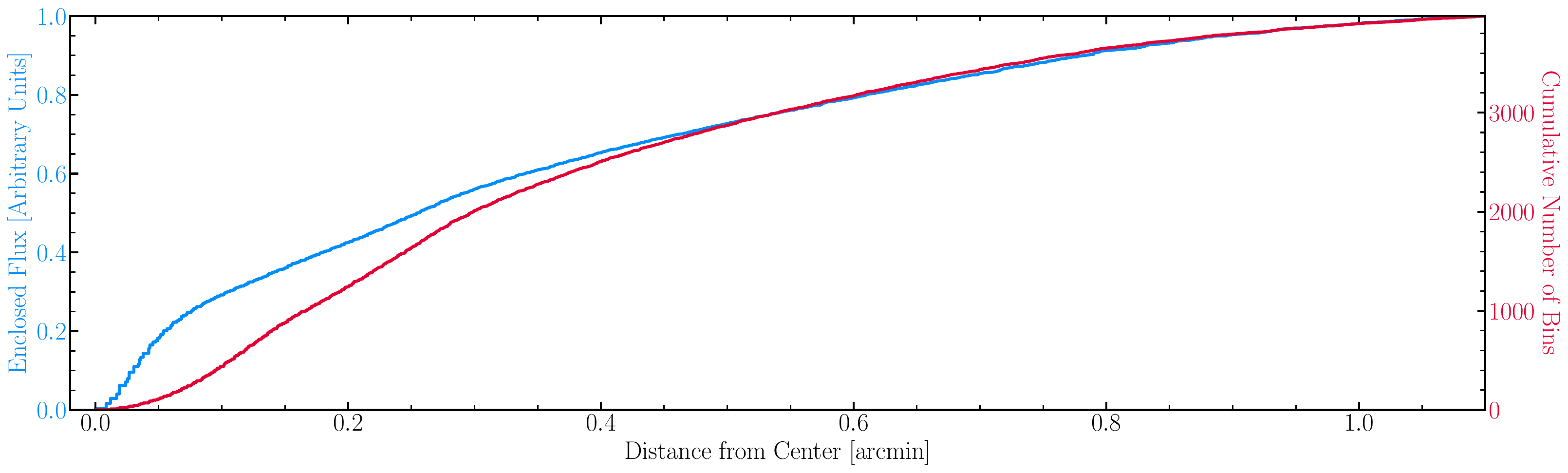}{\textwidth}{(d) Radial Profile}
    }
    \caption{The final binning of the FP data of NGC\,1068 generated from the \texttt{VorBin} routine. Panel (a) depicts the Voronoi tessellation map of NGC\,1068. The size of the Voronoi bins are chosen to achieve bins of a minimum intensity. Note how the bin size increases with distance from the center, as the intensity from the galaxy falls off. {Panels (b) and (c) present the number of pixels and SNR in a given bin respectively. Finally, panel (d) shows how the total number of bins changes with distance from the center (red), along with the total enclosed intensity at a given radius (blue)}. \label{fig:voronoi}}
\end{figure*}

\begin{figure*}[ht!]
    \centering
    \gridline{
        \fig{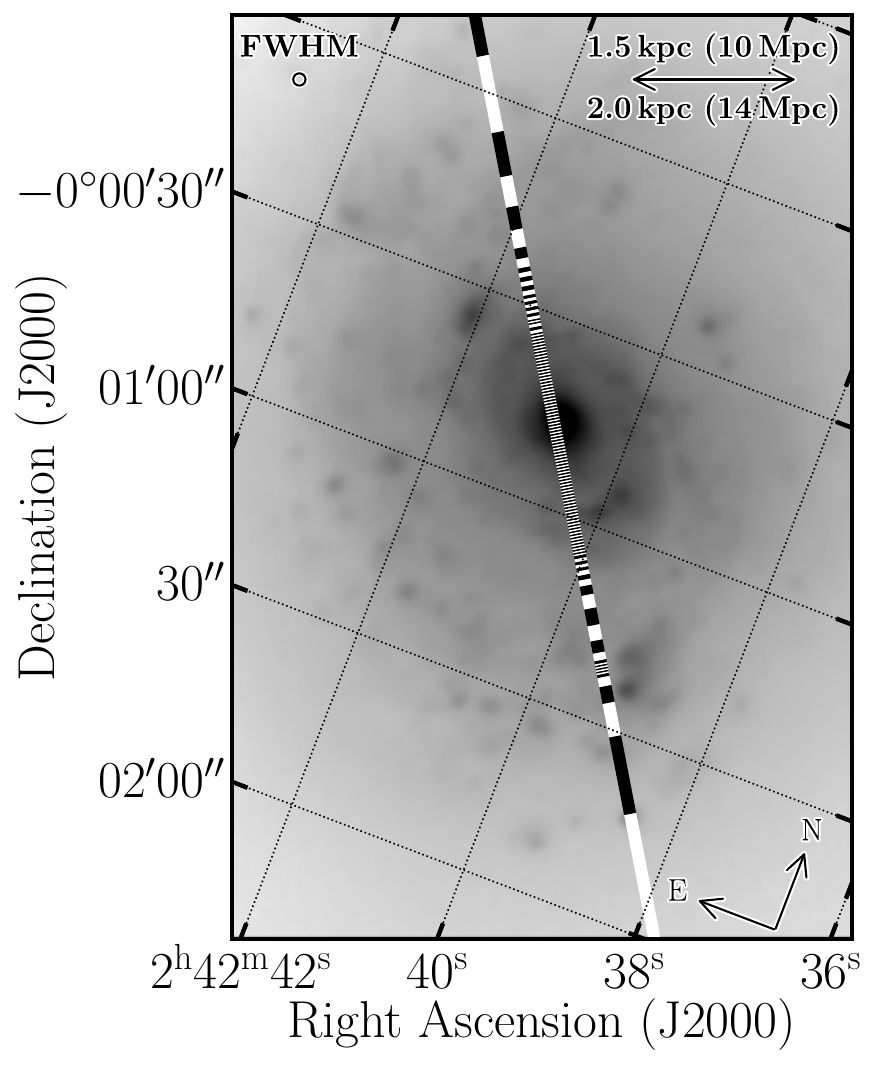}{0.28\textwidth}{(a) Voronoi Tessalation}
        \fig{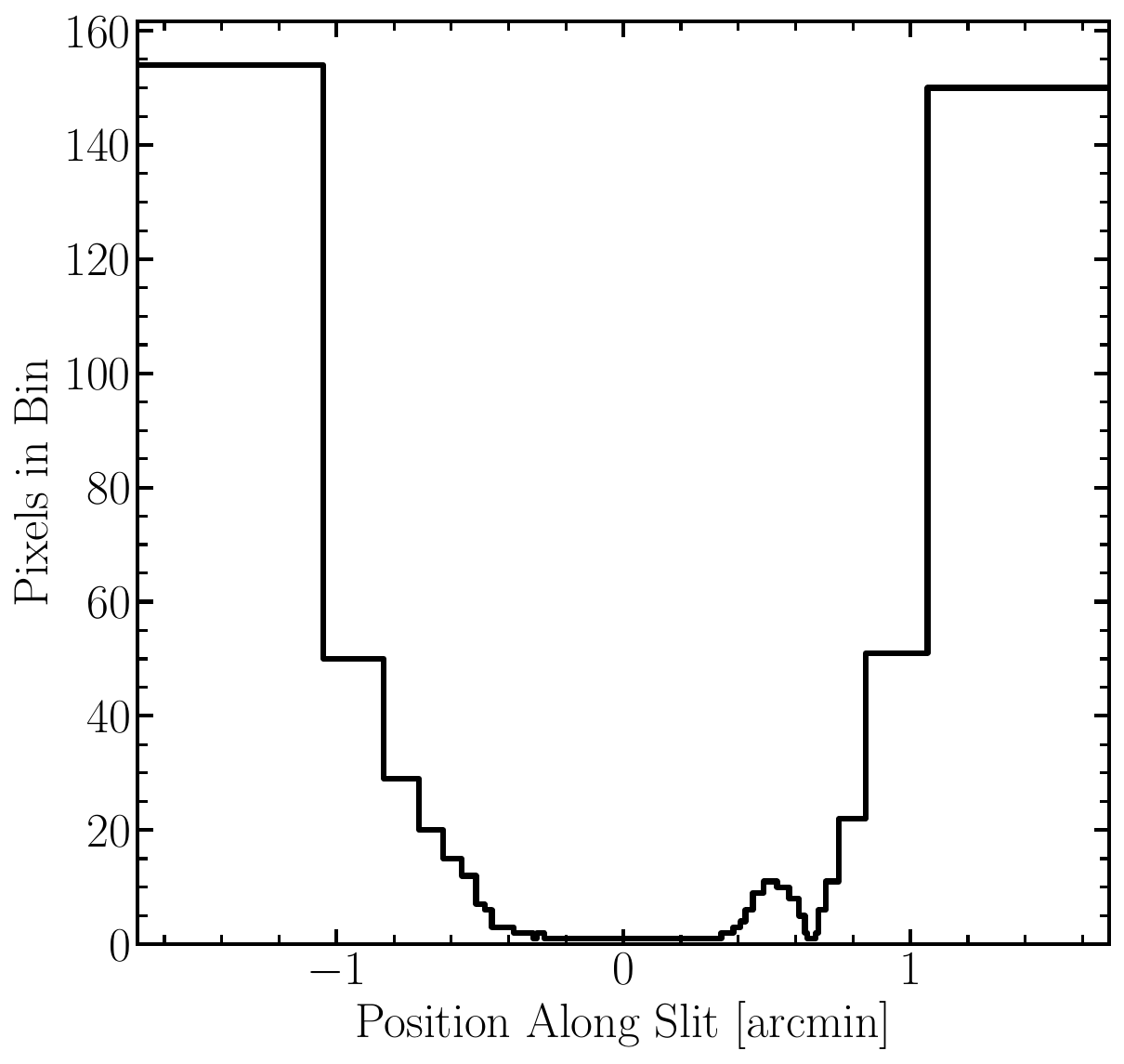}{0.36\textwidth}{(b) Voronoi Number}
        \fig{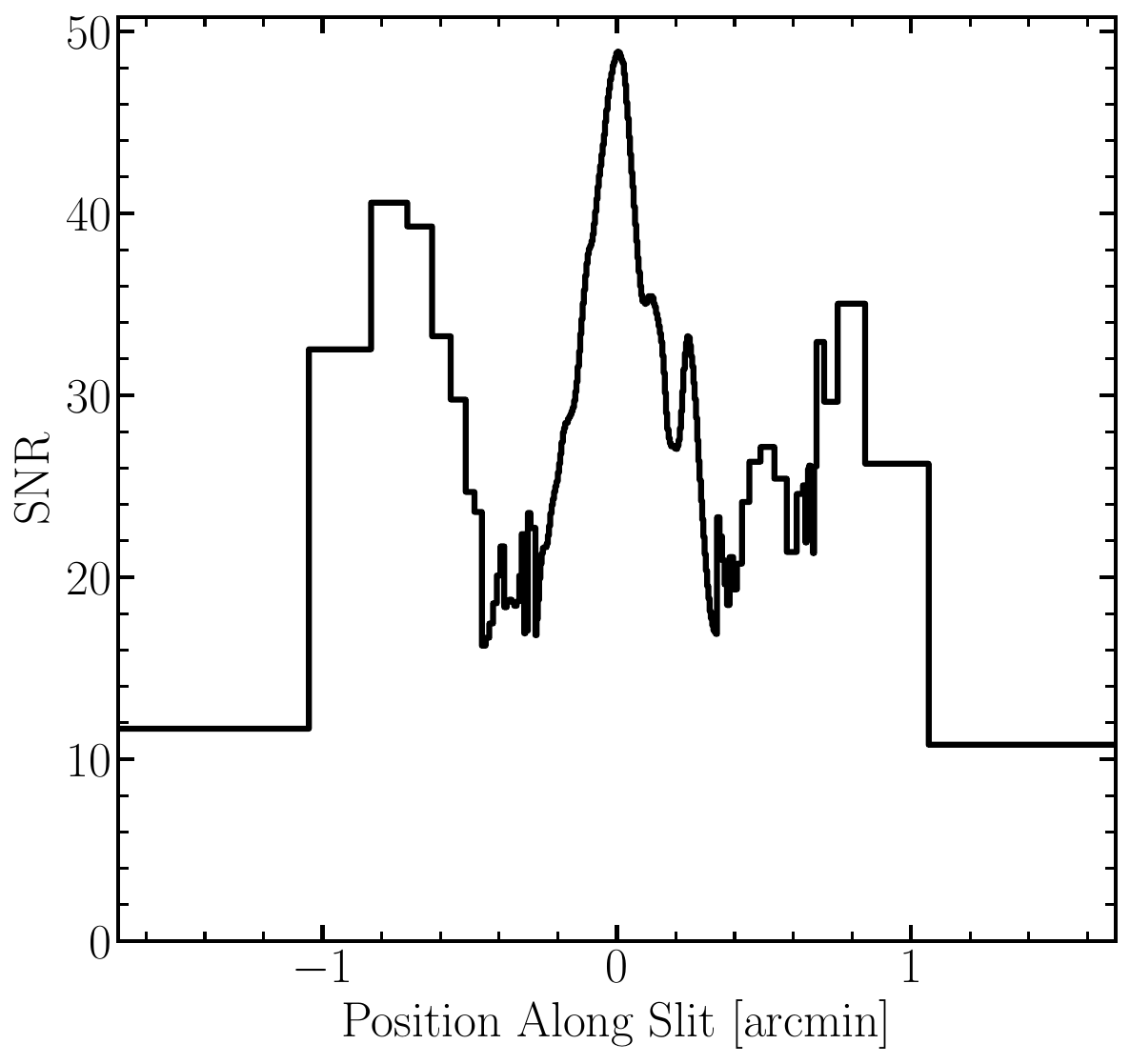}{0.36\textwidth}{(c) Voronoi SNR}
    }
    \gridline{
            \fig{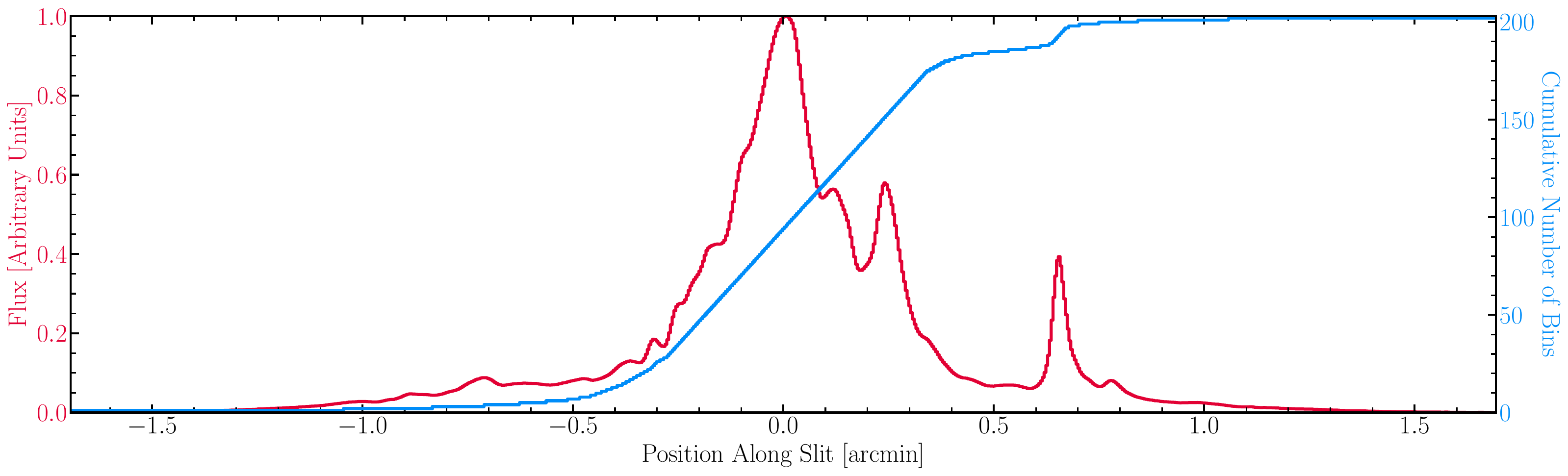}{\textwidth}{(d) Profile}
    }
    \caption{The final binning of the longslit data of NGC\,1068 generated from the \texttt{VorBin} routine. Panel (a) depicts the bin map of the longslit data. The size of the Voronoi bins are chosen to achieve bins of a minimum intensity. {Panels (b) and (c) present the number of pixels and SNR in a given bine respectively. Finally, panel (d) depicts the total intensity summed along the spectral dimension (red), along with the cumulative number of bins (blue)}.  \label{fig:longslitbin}}
\end{figure*}

\subsection{Adaptive Binning}
\label{subsec:adaptive}

We pursue an adaptive binning technique in order to achieve a minimum intensity in each bin to reach a sufficient signal to noise (SNR) far from the center of the galaxy. Adaptive binning creates nonuniform bins across the FoV, placing smaller bins in bright regions, and large bins in dim areas. For the RSS FP data we  pursue using a Voronoi Tessellation Algorithm, which produces a set of polygonal bins (a Voronoi diagram) such that the spaxels in each bin meets a minimum intensity or signal to noise threshold. We use the Voronoi binning algorithm, \texttt{VorBin}, by \citet{cappellariAdaptiveSpatialBinning2003} developed for IFU spectroscopy. 

{For both datasets we compute the SNR at each spatial position as the median SNR for all pixels across all wavelengths. For the longslit data we set a target SNR of 20 in order to ensure we can accurately retrieve emission-line fluxes. Due to the complex nature of the FP reduction and as our FP uncertainties are derived from the detector properties and not from combining multiple frames, as with our longslit data, we therefore set a more stringent target SNR of 100. We note that this does not guarantee a similar recovered SNR especially due to the small spectral range and relatively fewer data points across the observations.}

We show our Voronoi tessellation for NGC\,1068 in Figure \ref{fig:voronoi}. For the RSS Longslit data, we use VorBin to bin only along the spatial direction. Our final binning for the longslit data is presented in Figure \ref{fig:longslitbin}.

\subsection{Fitting Emission Lines}
\label{sec:fitemiss}

In order to measure the strength of nebular emission lines and the source of the ionization, we pursue fitting the emission lines required for the established \citet[][herafter BPT]{baldwinClassificationParametersEmissionline1981} diagnostic. 

\subsubsection{Fabry-P\'erot}

Following the Voronoi tessellation, we fit the [\ion{N}{2}]+H$\alpha$ complex in each voronoi bin. First, we take the set of spaxels from each dither in a given bin. The spaxels from a given dither are then normalized to the mean intensity of the set of spaxels from both dithers, accounting for the moving pupil of SALT. 

The [\ion{N}{2}]+H$\alpha$ complex is fitted with the sum of three Gaussians and a constant continuum offset. The redshifts of the two [\ion{N}{2}] lines are fixed to be equal, and the [\ion{N}{2}]$\lambda$6548\AA\ to [\ion{N}{2}]$\lambda$6583\AA\ flux ratio is fixed at 0.34 \citep{ohImprovedQualityassessedEmission2011}. {In addition, we require the redshifts of each of the lines to be bounded within 300\,km\,s$^{-1}$ of the systemic redshift of NGC\,1068 and that the flux in each line be positive. The FWHM of the lines is restricted to lie between 400\,km\,s$^{-1}$, consistent with our spectral resolution, and 1000\,km\,s$^{-1}$, the upper limit of narrow-line velocity widths.} Finally, the width of all of the lines are set to be the same as the limited spectral resolution does not allow us to distinguish line broadening due to physical effects. 

{We note that the underlying line-spread function (LSF) induced by the FP system is a Voigt profile \citep{rangwalaImagingFABRYPEROTSystem2008}. However this is primarily important for the wings of the emission lines and has a minimal effect on our ability to accurately measure the line height and center. In addition, as the widths of our emission lines are fixed to the same value for the FP fitting, our retrieved fluxes are not affected by the small difference between a Gaussian and the LSF. To validate this, we re-fit our data using Lorentzian profiles, a worst-case scenario for the differences between the assumed Gaussian profile and the true Voigt LSF, and find no noticeable differences to our fitting with Gaussian profiles. While an investigation of the kinematics of the emitting gas would require detailed knowledge of the underlying LSF, especially if paired with a higher-resolution etalon, in this work our Gaussian treatment of the lines paired with our $\Delta\lambda/2$ sampling is appropriate to retrieving emission line fluxes over the FOV.}

{The fitting is conducted using the \texttt{astropy} model fitting subclass using a Levenberg-Marquardt Least-Squares fitter. In order to generate errors on our recovered parameters, we bootstrap over the pixels in a given bin and refit 1000 times. The median and standard deviation of each parameter is taken as the value and error on the parameter. Finally we compute in which bins the median redshift was within 20\,km\,s$^{-1}$ of the limits mentioned above, while for the dispersion we calculate which bins are within 1\,km\,s$^{-1}$ of the limits. As the fitting in these bins was constantly hitting the limits in our fitting, the fits and their derived uncertainties cannot be trusted and are considered as failed fits. The subsequent analysis will only consider spaxels in which the fitting was successful which span a total angular size of 2.6\,arcmin$^2$.}

\subsubsection{Longslit}

Following the adaptive binning, we combine the spectra in each bin by summing their flux and adding their errors in quadrature. {We then fit the combined spectrum in each bin using the Galaxy/AGN Emission Line Analysis Tool \citep[GELATO,][]{GELATOv2.5.2} following the procedure outlined in \citet{hvidingNewInfraredCriterion2022}. GELATO models the continuum as a linear combination of Simple Stellar Populations (SSPs) from the Extended MILES stellar library \citep[E-MILES;][]{vazdekisUVextendedEMILESStellar2016} and models emission lines as Gaussians.}

{GELATO is a flexible Python framework that enables the fast and robust analysis of optical galaxy spectroscopy while specifying the relationship between various emission line parameters to suit the user's needs. For the [\ion{N}{2}] and [\ion{O}{3}] emission doublets we tie the respective velocity dispersions and redshifts together and set the emission line ratios, [\ion{N}{2}]$\lambda$6548\AA/[\ion{N}{2}]$\lambda$6583\AA\ and [\ion{O}{3}]$\lambda$5007\AA/[\ion{0}{3}]$\lambda$4959\AA, to 0.34 and 0.33 respectively 
\citep{ohImprovedQualityassessedEmission2011}. In addition, we do not attempt to fit a broad-line component to the Balmer emission lines.}

\section{Results}
\label{sec:results}

In this section we present our maps of NGC\,1068 generated from fitting the emission lines from the reduced FP and longslit data. 

\begin{figure*}[ht!]
    \centering
    \includegraphics[width=\textwidth]{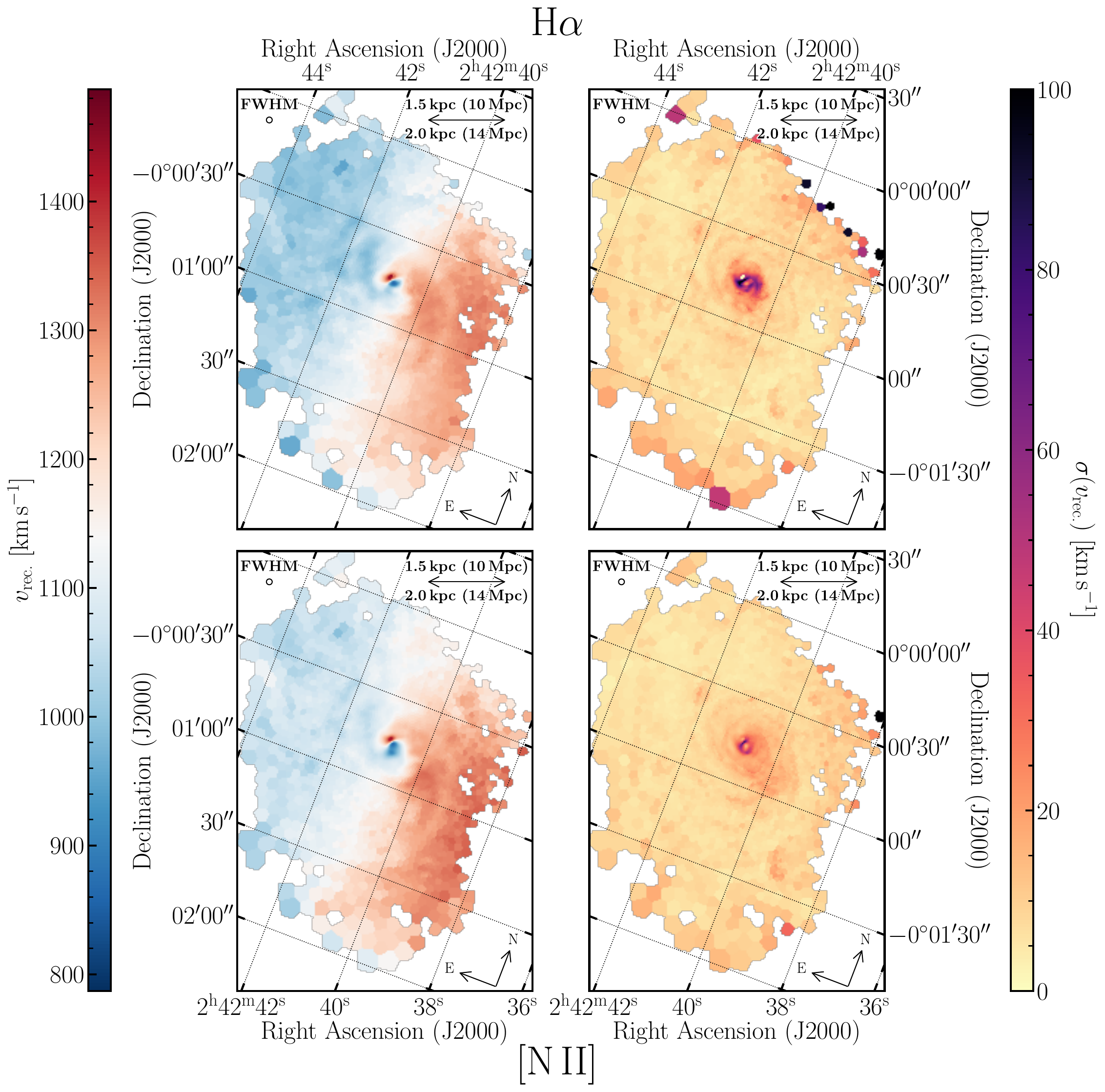}
    \caption{Velocity maps (left) {and their associated standard deviations (left)} for the H$\alpha$ (top) and [\ion{N}{2}] (bottom) emission lines. The colorbars for the velocity are centered on the systemic redshift of NGC\,1068 corresponding to 1137\,km\,s$^{-1}$. \label{fig:vel}}
\end{figure*}

\begin{figure*}[ht!]
    \includegraphics[width=\textwidth]{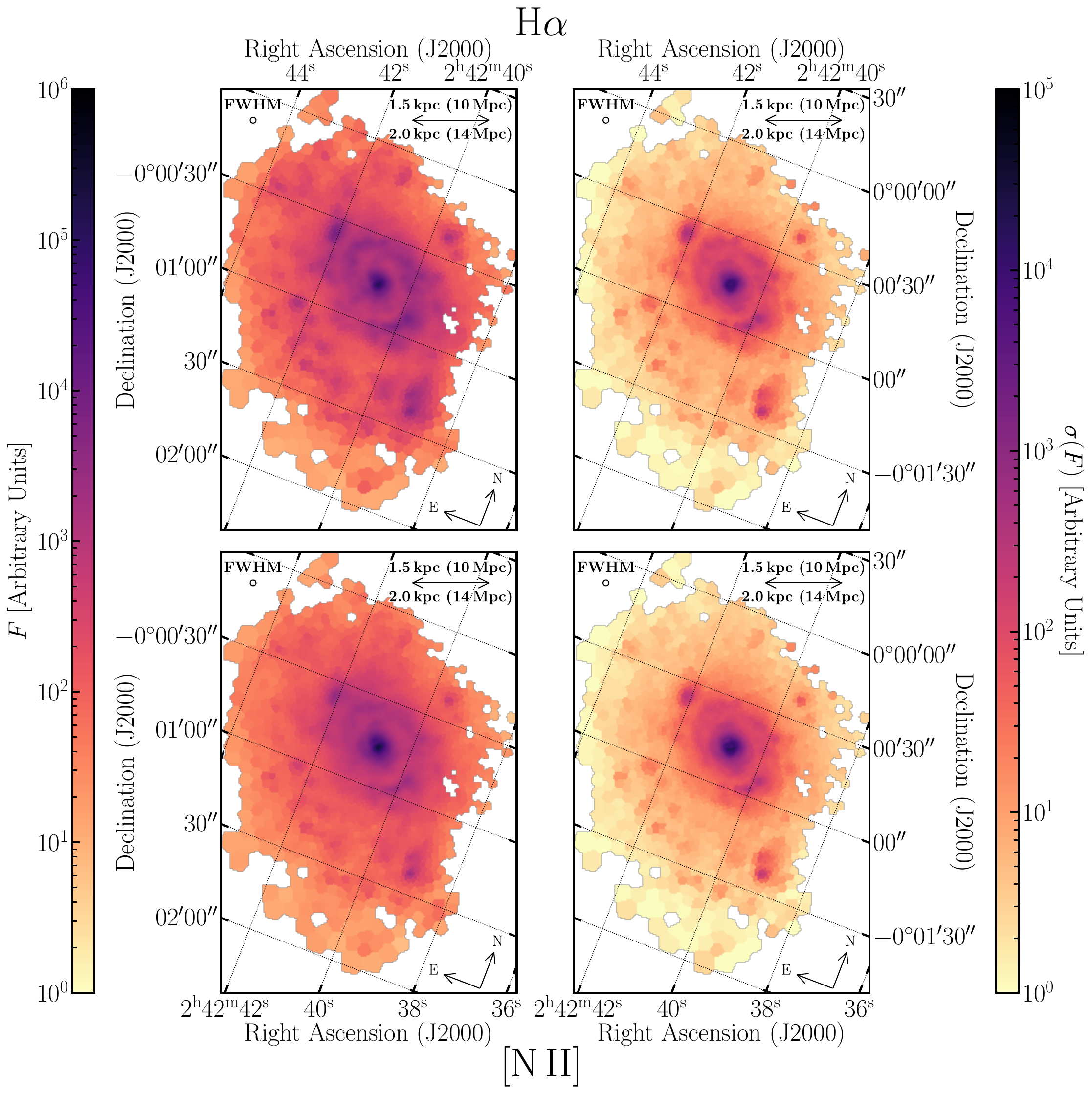}
    \caption{Emission line flux maps (left) {and their associated standard deviations (right)} for the H$\alpha$ (top) and [\ion{N}{2}]$\lambda6583$\AA\ (bottom) lines. While there is no absolute flux calibration due to the nature of the stationary primary mirror of SALT, the relative flux is preserved. \label{fig:flux}}
\end{figure*}

\begin{figure*}[ht!]
    \includegraphics[width=\textwidth]{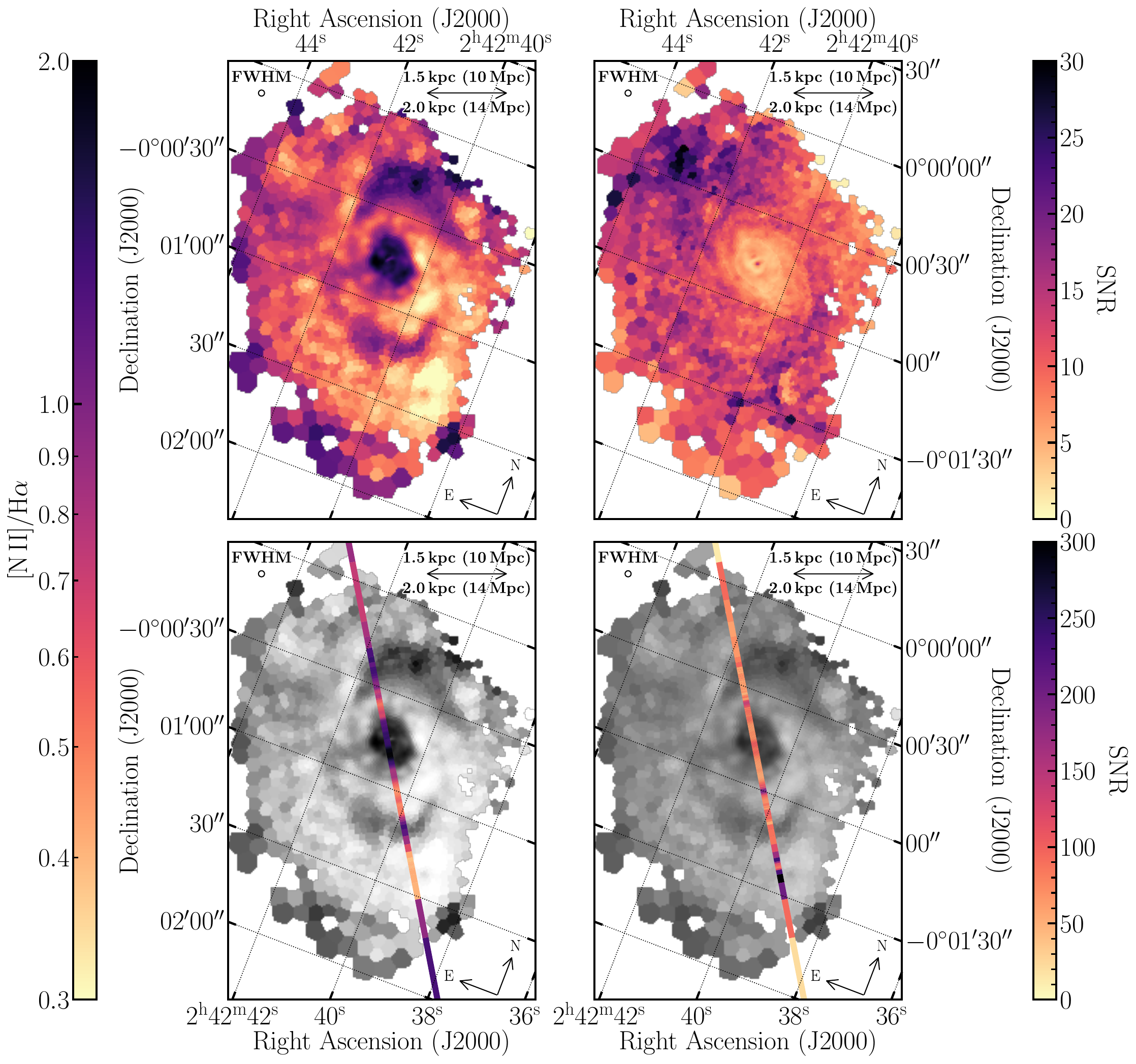}
    \caption{Maps of H$\alpha$/[\ion{N}{2}] (left) {and their associated SNRs (right)} for the FP (top) and longslit (bottom) data. The longslit observations are plotted over grayscale FP H$\alpha$/[\ion{N}{2}] ratio maps. {Their corresponding standard deviations are shown in panels (c) and (d).} The longslit observations confirm the accuracy of the our FP data reduction and analysis and highlight the kiloparsec scale ionizing features. \label{fig:ion}}
\end{figure*}

\subsection{Velocity Maps}

We present maps of the recessional velocity for both the H$\alpha$ and [\ion{N}{2}] emission lines in NGC\,1068 {and their associated errors in} Figure \ref{fig:vel}. The rotational motion of the galaxy is apparent in the figure, with the westward side receding from the observer relative to the eastward side. Nearer towards the center of the galaxy, the change in the inclination of the disk becomes apparent as observed in \citet{schinnererBarsWarpsTraced2000} by tracing the molecular gas in the inner 20\arcsec\ of the galaxy and attributed to a warped disk in NGC\,1068. We are able to observe the same warp in the ionized gas and trace the inclination angle of the disk out to several arcminutes.

\begin{figure*}[ht!]
    \centering
    \gridline{
        \fig{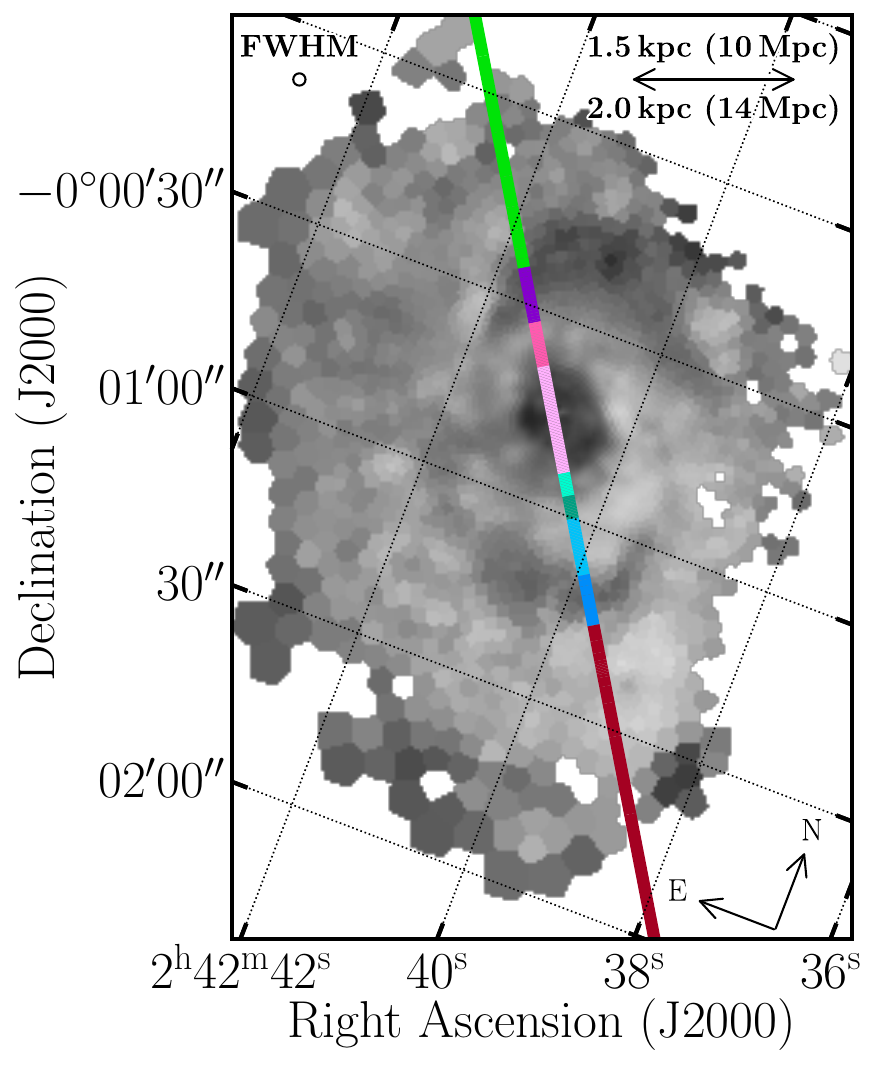}{0.43\textwidth}{(a) Longslit Position}
        \fig{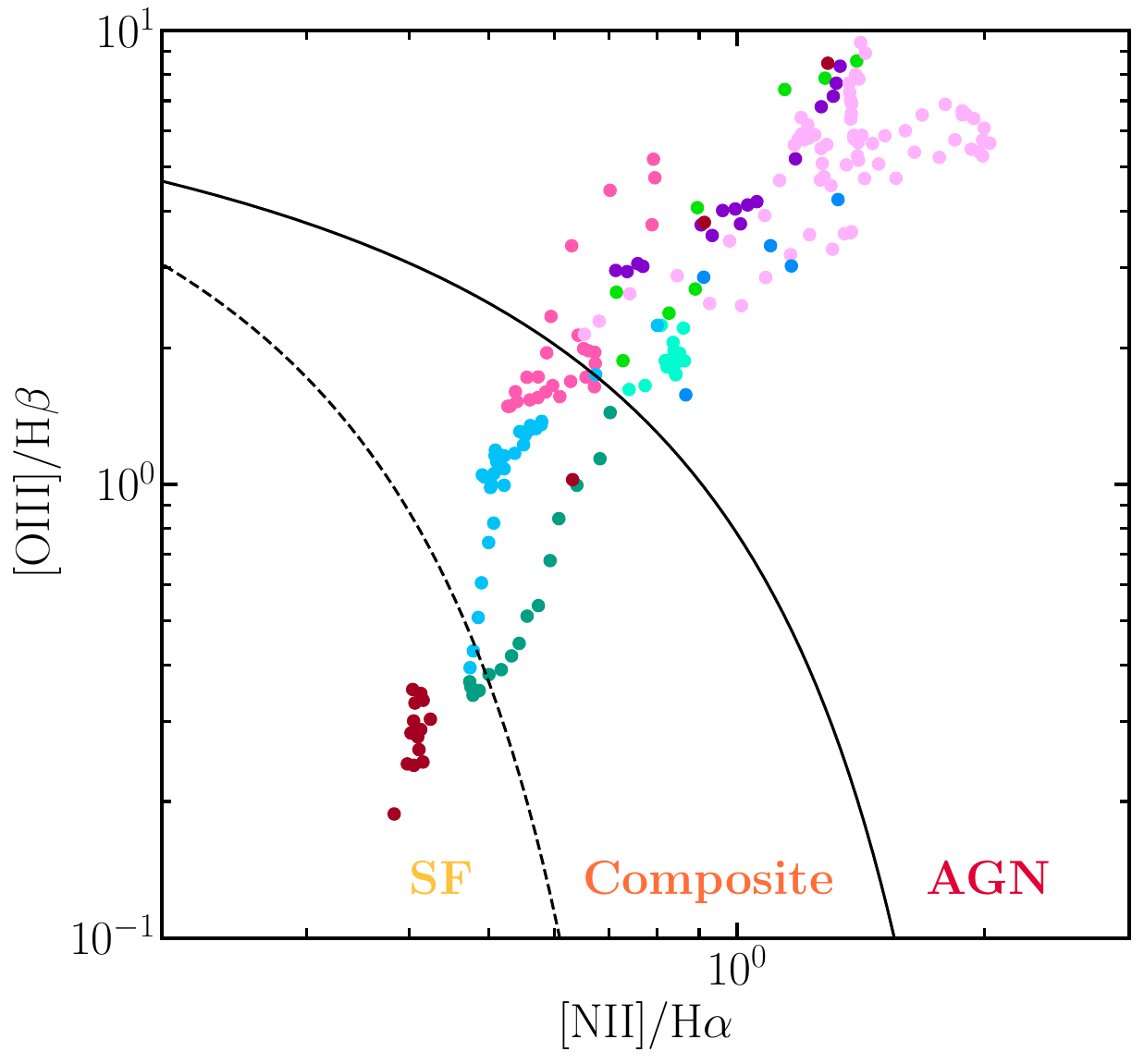}{0.57\textwidth}{(b) BPT Position}
    }
    \caption{In panel (a), colored longslit positions overplotted on the FP $\log_{10}$(H$\alpha$/[\ion{N}{2}]) diagnostic map. In panel (b) we plot BPT positions of the colored longslit positions to highlight the transition between the star-forming and AGN regions of NGC\,1068 along the longslit spectrum. With the additional [\ion{O}{3}] and H$\beta$ information afforded by the longslit spectrum, the source of the kiloparsec scale ionization features becomes clear as the line ratios are consistent with that of an AGN. In addition, we plot the \citet{kewleyTheoreticalModelingStarburst2001} and \citet{kauffmannStellarMassesStar2003} demarcations as solid and dashed lines, respectively.\label{fig:bpt}}
\end{figure*}

\subsection{Emission Line Maps}

We present maps of the emission line fluxes for both [\ion{N}{2}] and H$\alpha$ in NGC\,1068 {along with their associated errors} in Figure \ref{fig:flux}. Both emission line maps clearly reveal the ring of star formation around the the galaxy center, as documented in \citet{thronsonNearinfraredImageNGC1989}, which are more prominent in H$\alpha$. Areas of star formation throughout the disk become apparent as well. 

\subsection{Ionization Map}

In order to assess the ionization state across the FoV, we produce maps of the [\ion{N}{2}]$\lambda$6583\AA/H$\alpha$ emission line ratio. This diagnostic is sensitive to the level of ionization in the gas and can be used to discern the source of ionizing radiation when combined with other emission line ratios \citep{baldwinClassificationParametersEmissionline1981}. Since the relative fluxes are preserved in the FP maps, the line ratio is measured accurately even without absolute flux calibration. While this emission line ratio alone does not totally discern the ionization source of the gas, it gives a measurement of the level of ionization. Our ionization map, presented in Figure \ref{fig:ion} {along with it's associated SNR}, reveals the kiloparsec scale high-ionization features previously observed by \citet{dagostinoStarburstAGNMixingTYPHOON2018}. 

In order to verify our FP ionization map, we overlay the longslit line ratios over the FP map again in Figure \ref{fig:ion} {along with it's associated SNR}. The previous longslit data obtained of NGC\,1068 covers the extended high-ionization regions {and is offset from the center of the kiloparsec-scale ionized features by $\sim$20$^\circ$}. The longslit results verify the existence of the highly ionized features in our FP data. {We note that at the edges of the longslit data we observe some disagreement with our FP measurements, especially in the southwestern region of our data, though we note that this is likely driven by the low SNR measurements in both datases at those distances.}

{We note, however, that the [\ion{N}{2}]/H$\alpha$ ratio is also sensitive to the gas-phase metallicity and is therefore an imperfect tracer of the ionization ratio. Without the availability of [\ion{O}{3}] and H$\beta$ FP maps we are unable to break this degeneracy over the entire field of view.} However, as the longslit spectrum has wavelength coverage of the [\ion{O}{3}]+H$\beta$ complex, we plot the BPT position of the longslit spectrum bins overplotted on the FP ionization map in Figure \ref{fig:bpt}. In addition, we plot the \citet{kauffmannStellarMassesStar2003} and \citet{kewleyTheoreticalModelingStarburst2001} demarcations on the BPT diagram. The longsit spectrum observations confirm that the ionized features have line ratios consistent with an AGN ionizing source. The transition along the longslit FoV becomes apparent, with the kiloparsec scale ionized features having AGN line ratios as strong as the central engine, but the intermediate regions having composite and star formation emission ratios, diminished.

\begin{figure*}[ht!]
    \gridline{
        \fig{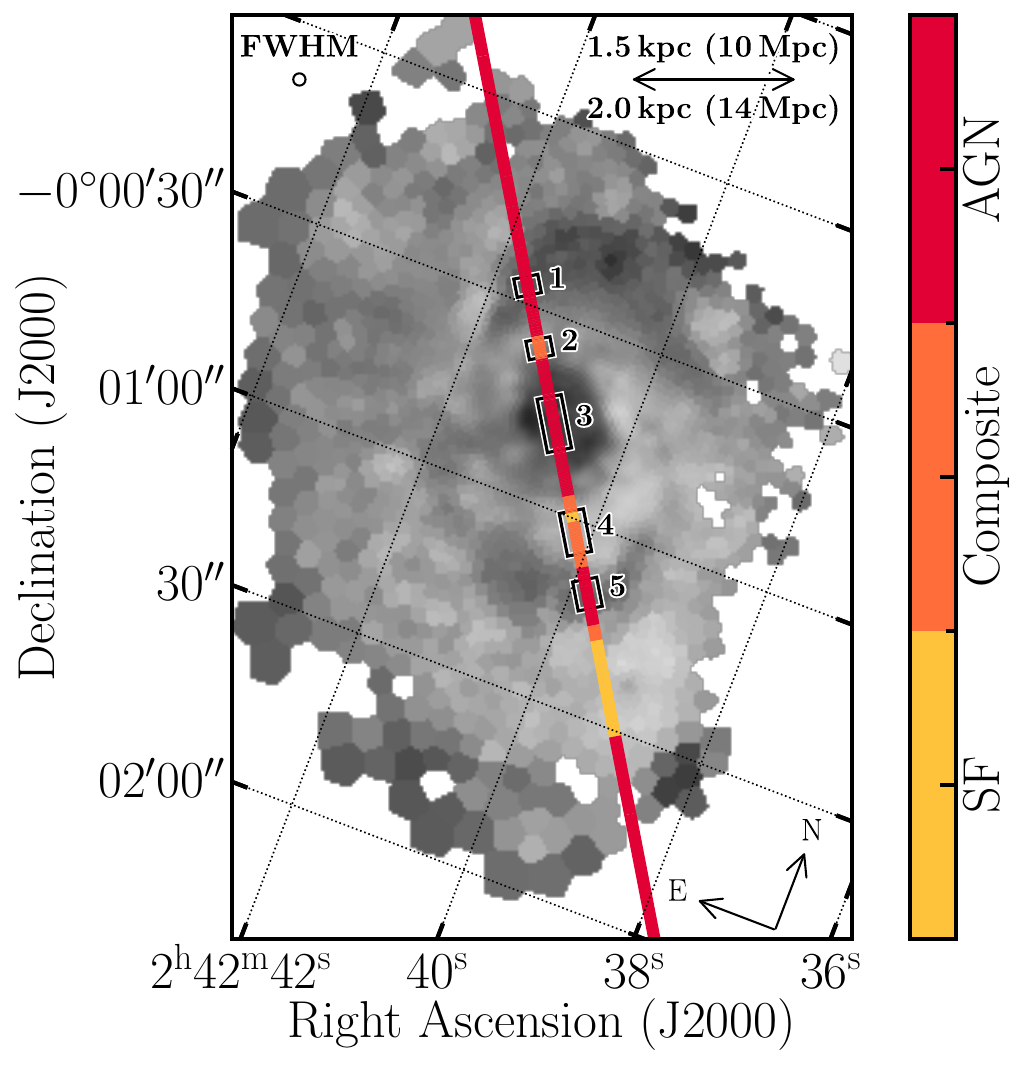}{0.43\textwidth}{(a) Longslit Position}
        \fig{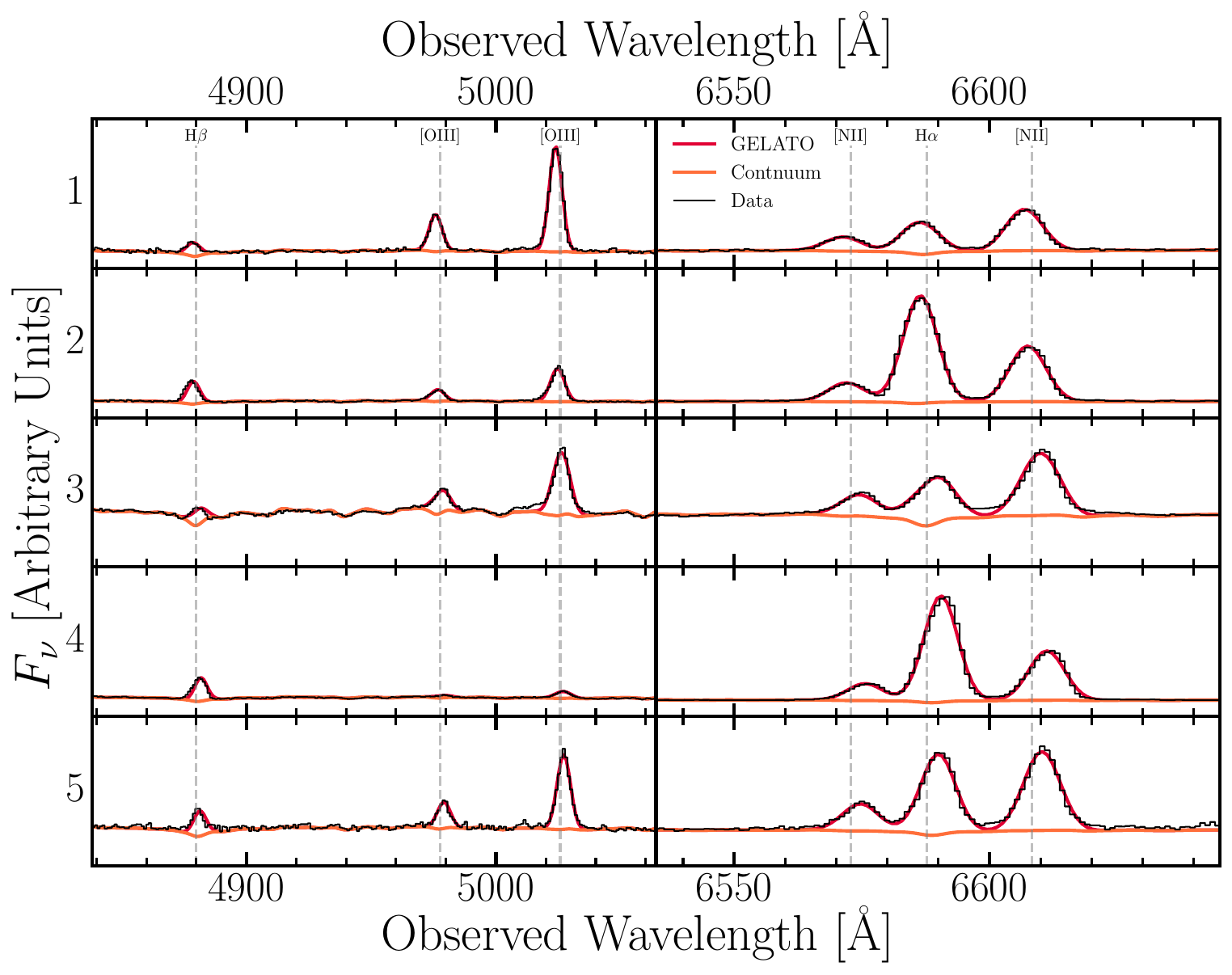}{0.57\textwidth}{(b) Stacked Spectra }
    }
    \caption{In panel (a), the BPT positions of the longslit spectrum bins overplotted on the FP $\log_{10}$(H$\alpha$/[\ion{N}{2}]) diagnostic map. In panel (b), we present the stacked spectra from the highlighted regions in panel (a). The selected regions highlight the transition of the longslit spectrum from star-forming to AGN activity across the FoV. In addition, the spectra reveal the rotational motion of the galaxy with the slight shift of the emission lines relative to the average redshift of the galaxy. \label{fig:spectra}}
\end{figure*}

To emphasize this transition further, we plot the BPT position across the longslit FoV alongside stacked spectra across from the central AGN dominated region, the kpc-scale ionized regions, and the transition regions between the two in Figure \ref{fig:spectra}. {The regions over which the spectra are stacked are chosen by eye and to emphasize the differences across the FoV.} The stacked spectra show a clear transition from AGN dominated, with prominent [\ion{O}{3}] and [\ion{N}{2}] in the central region, to composite and star forming, back to AGN dominated in the kiloparsec scale features. We note that the the indication of AGN regions beyond the northern and southern extended features is due to a lack of signal in the longslit image, and likely not indicative of AGN activity out to the edge of the figure.

{Figure \ref{fig:spectra} also presents the GELATO fits to the stacked spectra along with the retrieved continuum to highlight the strength of absorption features in particular below the H$\alpha$ emission line. It becomes clear that across the galaxy the emission line strength dominates over and the absorption features negligibly contribute to the measurement of the H$\alpha$ flux, except for perhaps at the center of the galaxy. This validates our fitting approach to the FP data, where we are not able to model the underlying continuum, and reinforces that our FP ionization measurements are accurate over the FoV.}

\subsection{Potential Light Echoes}
\label{subsec:lightecho}

{The kilo-parsec scale features visible in Figure \ref{fig:ion} are attributed to AGN ionization by \citet{dagostinoStarburstAGNMixingTYPHOON2018}. This picture is reinforced by Figure 29 of \citet{dagostinoStarburstAGNMixingTYPHOON2018} which shows no enhancement of the velocity dispersion in the location of the shocks. In addition, if shocks driven by outflows were present, we would expect to see distortions in the velocity field of the galaxy or enhanced line widths in the kiloparsec scale ionized regions \citep{hoSAMIGalaxySurvey2014}. As the ionization features are nearly perpendicular to the axis about which the galaxy is inclined, these velocity shifts would be visible in the velocity field of the galaxy.}

We do not observe evidence of an outflow in the velocity fields of the H$\alpha$ or [\ion{N}{2}] as shown in Figure \ref{fig:vel}. In addition, the kiloparsec scale ionized regions have velocity dispersions consistent with the intervening composite and star forming regions, with average velocity dispersions of $\sim$175\,km\,s$^{-1}$ with a maximum measured dispersion of $\sim$225\,km\,s$^{-1}$, while the central AGN region does show enhanced velocity dispersions, with average velocity dispersions of $\sim$325\,km\,s$^{-1}$ with a maximum measured dispersion of $\sim$450\,km\,s$^{-1}$. 

{In this section we explore the possibility of modeling the kiloparsec-scale AGN-photoionized emission across the galaxy to investigate if the spatial variation can provide insight into the AGN luminosity as a function of time. We present the possibility that the ionized features are due to past AGN activity and are therefore a form of light echoes, and create a toy model to explore the time variability of NGC\,1068.}

\subsubsection{AGN Contribution to H$\alpha$}

{Similar to previous spatially-resolved spectroscopic studies of AGN, we aim to disentangle the contribution of AGN and SF processes to the H$\alpha$ flux across the FoV \citep{kewleyOpticalClassificationSouthern2001,daviesDissectingGalaxiesSpatial2016,daviesDissectingGalaxiesSeparating2017,dagostinoStarburstAGNMixingTYPHOON2018,dagostinoNewDiagnosticSeparate2019}.}
We first derive the relative H$\alpha$ flux contributed by the AGN in the FP image. By assigning typical star forming and AGN [\ion{N}{2}]/H$\alpha$ ratios, we can solve for the AGN contribution given the H$\alpha$ and [\ion{N}{2}] fluxes. By examining the range of [\ion{N}{2}]/H$\alpha$ ratios present in the longslit data we can take the extreme values to be representative of pure SF or AGN activity. By investigating Figure \ref{fig:bpt}, we set the pure-AGN value as {$({\text{[\ion{N}{2}]/H}}\alpha)_{\text{AGN}} = 2$ and a the pure star-forming value as $({\text{[\ion{N}{2}]/H}}\alpha)_{\text{SF}} = 0.3$}. We can therefore express the relationships between the line ratios and the typical emission ratio values in the following equations:
\begin{equation}
    {\text{H}}\alpha = {\text{H}}\alpha_{\text{AGN}} + {\text{H}}\alpha_{\text{SF}}
\end{equation}
\begin{equation}
    {\text{[\ion{N}{2}]}} = {\text{[\ion{N}{2}]}}_{\text{AGN}} + {\text{[\ion{N}{2}]}}_{\text{SF}}
\end{equation}
\begin{equation}
    {\left({\text{[\ion{N}{2}]}}/{\text{H}}\alpha\right)_{\text{AGN}}\times{\text{H}}\alpha_{\text{AGN}} = {\text{[\ion{N}{2}]}}_{\text{AGN}}}
\end{equation}
\begin{equation}
    {\left({\text{[\ion{N}{2}]}}/{\text{H}}\alpha\right)_{\text{SF}}\times{\text{H}}\alpha_{\text{SF}} = {\text{[\ion{N}{2}]}}_{\text{SF}}}
\end{equation}
By solving the above system of equations, we can derive the H$\alpha$ flux contributed to the AGN using the H$\alpha$ and [\ion{N}{2}] emission maps and the typical line ratios chosen above.
\begin{equation}
    {{\text{H}}\alpha_{\text{AGN}} = \frac{\left({\text{H}}\alpha/{\text{[\ion{N}{2}]}}\right)_{\text{SF}}\times{\text{[\ion{N}{2}]}}- {\text{H}}\alpha}{\left({\text{[\ion{N}{2}]}}/{\text{H}}\alpha\right)_{\text{AGN}}\times\left({\text{H}}\alpha/{\text{[\ion{N}{2}]}}\right)_{\text{SF}} - 1}}
\end{equation}
The derived map of the H$\alpha$ flux contributed by the AGN is shown in the {left middle} panel of Figure \ref{fig:radprof}. {The derived map demonstrates that there is significant variation over the FoV, especially in the direction of the kiloparsec-scale ionized features that may be attributable to varying AGN intensity in the past.}

\subsubsection{Toy Model}

In order to recreate the structures visible in the H$\alpha_{\text{AGN}}$ flux, we construct a toy model of the  NGC\,1068 galaxy and its AGN. The southern structure motivated our attempt to model the NGC\,1068 AGN system as a biconical outflow intersecting with a galactic disk. We model the disk of the galaxy as an opaque plane, consistent with measurements of the optical depth of galaxy disks \citep{jamesMeasurementOpticalDepth1993}. If we align the $x$ and $y$ axes with our observation (with the $z$ axis therefore towards the observer), we can start be defining the plane of the disk in the $x$ and $z$ axes, defined by the normal vector, $\hat{n} = \mathbf{\hat{j}} = (0,1,0)$. Given that a face on galaxy has an inclination angle of 0$^\circ$, therefore $\hat{n}$ defines a disk with an inclination of 90$^\circ$. To rotate the model disk to an arbitrary angle, $i_\textrm{gal}$, we apply the standard rotation matrix about the $x$-axis, $\textbf{R}_x$:
\begin{equation}
\hat{n}_\textrm{inclined}=
\textbf{R}_x\left(\frac{\pi}{2} - i_\textrm{gal}\right)\times
\mathbf{\hat{j}}
=\begin{pmatrix}
0\\
\sin(i_\textrm{gal})\\
\cos(i_\textrm{gal})
\end{pmatrix}
\end{equation}

We then apply a second rotation about the $z$-axis, $\textbf{R}_z$, to align the model with an arbitrary position angle, $\textrm{PA}_\textrm{gal}$. The position angle of the observation, determined from the WCS information, is added to the model galaxy PA in order to match with the data. Therefore, the normal vector that describes the modeled disk of NGC\,1068 is given by:
\begin{equation}
\hat{n}_\textrm{gal}=\textbf{R}_z(\textrm{PA}_\textrm{gal}+\textrm{PA}_\textrm{WCS})\times\textbf{R}_x(\pi/2 - i_\textrm{gal})\times\mathbf{\hat{j}}
\end{equation}

\begin{figure*}[ht!]
    \includegraphics[width=\textwidth]{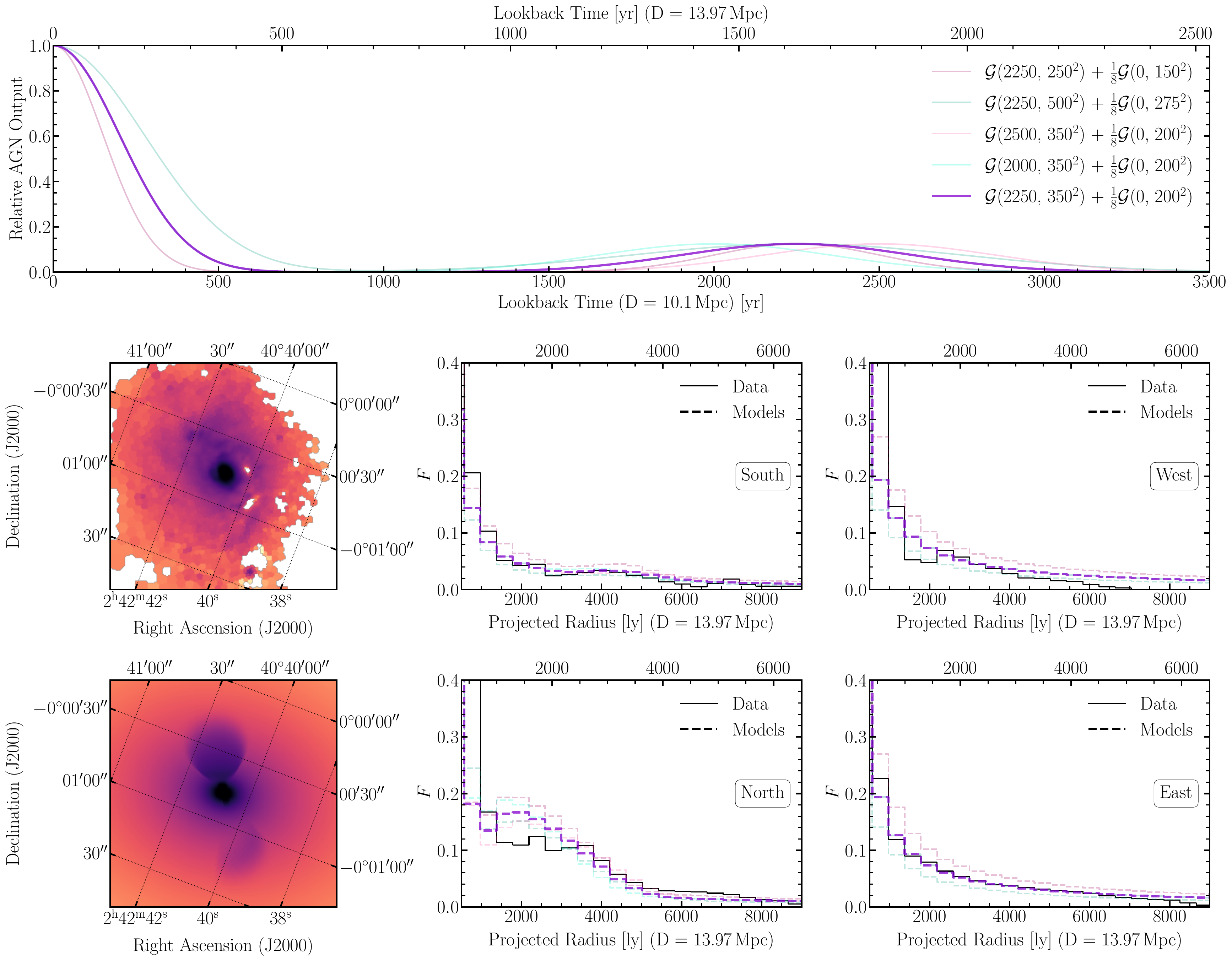}
    \caption{{Our model of the kiloparsec scale ionized region as light echoes from the AGN in NGC\,1068. 
    In the top panel we present the model lightcurves with the main model depicted in solid purple.
    The center left panel shows the derived (H$\alpha$)$_\textrm{AGN}$ flux in log scale. In the bottom left panel we show the modeled AGN flux following the procedure described in Section \ref{subsec:lightecho} in log scale. We plot the radial profiles of the derived (H$\alpha$)$_\textrm{AGN}$ flux (solid) and the model AGN lightcurves (dashed) for the four cardinal directions in the center middle (South), right middle (West), center bottom (North), and right bottom (East) panels. \label{fig:radprof}}}
\end{figure*}

% NGC\,1068 is inclined at an angle of roughly 40$^\circ$ with a position angle (PA) of 0-10$^\circ$ \citep{devaucouleursSouthernGalaxiesVI1973,brinksHIObservationsNGC1997}. 
We set the position angle of the disk to be 8$^\circ$ based on \citet{garcia-burilloALMAResolvesTorus2016}\footnote{{As our PA is defined as the angle to the normal vector of the galaxy rather than the axis about which the galaxy rotates, our PA is greater by 90$^\circ$.}} and the inclination to be 40$^\circ$ \citet{schinnererBarsWarpsTraced2000}. As we model the disk as an opaque surface, so any points behind the plane of the disk do not contribute any flux to the toy model:
\begin{equation}
F\left(\hat{n}_\textrm{gal}\cdot\begin{pmatrix}
x\\
y\\
z
\end{pmatrix} < 0\right) = 0
\label{eqn:disk}
\end{equation}

In order to match the ionization features, we model bursts of past AGN activity as a biconical structure. In general, the equation for the region inside the bicone along the $z$-axis with opening angle, $\theta_\textrm{o}$, is given by: 
\begin{equation}
    x^2 + z^2 \le \tan^2\left(\frac{\theta_\textrm{o}}{2}\right)y^2
\end{equation}
We can repeat the same process as above to generate the generic rotation matrix  for the disk of the galaxy to find the normal vector that defines the direction of the bicone:
\begin{equation}
\mathbf{R}_\textrm{c}=\textbf{R}_z(\textrm{PA}_\textrm{cone}+\textrm{PA}_\textrm{WCS})\times\textbf{R}_x(i_\textrm{cone})
\end{equation}
Therefore, we can apply the following constraint:
\begin{equation}
F(\vec{x'}) = 0
\label{eqn:cone}
\end{equation}
where $\vec{x'} = (x',y',z')$ and 
\begin{equation}
x'^2 + z'^2 > \tan^2\left(\frac{\theta_\textrm{o}}{2}\right)y'^2
\textrm{ where}\begin{pmatrix}
x' \\
y' \\
z'
\end{pmatrix}
=
\mathbf{R}_\textrm{c}
\begin{pmatrix}
x \\
y \\
z
\end{pmatrix}
\end{equation}

{We set the opening angle of the bicone to 80$^\circ$ based on the work done in \citet{dasKinematicsNarrowLineRegion2006} and \citet{garcia-burilloALMAResolvesTorus2016}. For the inclination angle, \citet{crenshawResolvedSpectroscopyNarrowLine2000} and \citet{dasDynamicsNarrowLineRegion2007} find a small inclination of $\sim$5$^\circ$ for the innermost 6'' of the galaxy. We find that to best fit our data we require a slight smaller inclination of $\sim$2.5$^\circ$ into the plane of the sky, but note that our observations are at a larger scale than the previous work. Finally, the position angle of the bicone is set to 30$^\circ$ based on \citet{dasKinematicsNarrowLineRegion2006}.}

{We model the bursts of AGN activity as two Gaussians in time: $\mathcal{G}(\mu,\sigma) = \exp(-\frac{1}{2}((t_\textrm{lookback} - \mu)/\sigma)^2)$. Assuming a distance of 10.1\,Mpc, our best fit models are comprised of a past burst occurring $\sim$2250 years ago and a current burst of activity beginning within the past $\sim$500 years that is 8 times in strength. The variable output of the AGN is shown in the top panel of Figure \ref{fig:radprof}. In addition, we present a few other potential models of AGN activity in Figure \ref{fig:radprof} on which the subsequent analysis is also performed in order to provide some context on the dependence on our chosen Gaussian parameters.} Finally we add a constant background AGN component that emits along the plane of the disk and is not restricted to the biconical structure {which is intended to represent a constant level of photoionization induced by the AGN in the galaxy.}

In order to compare our results to our FP maps, we simulate observations of our model. We create a $\sim$6\,kpc wide cube grid with a resolution that matches our pixel scale (80\,ly\,px$^{-1}$) and is aligned the FP image. First, we calculate the distance to each point: $d(\vec{x}) = \sqrt{x^2+y^2+z^2}$, where $\vec{x}$ is the position vector to an arbitrary grid cell. In order to realistically model the travel time from the central engine to model position and finally to the observer, the observed flux is given by: 
\begin{equation}
F_\textrm{AGN}(\vec{x}) \propto \frac{1}{d(\vec{x})^2}L_\textrm{AGN}\left(\frac{d(\vec{x}) - x}{c}\right)
\end{equation}
where $c$ is the speed of light. In order to account for the gas in the galaxy, above the opaque plane, we model the disk as an exponential fall of with a scale height of $10^{3.5}$\,ly ($\sim1$\,kpc):
\begin{equation}
F_\textrm{AGN}(\vec{x}) \propto \frac{\exp(\vec{x}\cdot\hat{n}_\textrm{gal}/10^{3.5})}{d(\vec{x})^2}L_\textrm{AGN}\left(\frac{d(\vec{x}) - x}{c}\right)
\end{equation}

After applying the constraints outlined in Equations \ref{eqn:disk} and  \ref{eqn:cone}, we create a simulated image by summing the model flux grid along the line of sight to the observer: $F_\textrm{im}(x,y) = \sum_z F(\vec{x})_\textrm{AGN}$. Finally, a constant AGN background component in the disk is modeled by calculating the distance to a point in the disk at each position in the simulated image: 
\begin{equation}
    d_\textrm{proj}(x,y) = x^2 + y^2 + \left(\frac{x(\hat{n}_\textrm{gal})_x + y(\hat{n}_\textrm{gal})_y}{(\hat{n}_\textrm{gal})_z}\right)^2
\end{equation}
The background flux is then given by:
\begin{equation}
    F_b(x,y) \propto \frac{1}{(d_\textrm{proj}(x,y))^2}
\end{equation}

The final simulated image, calculated by summing the AGN and background components, is presented in the {left bottom} panel of Figure \ref{fig:radprof} along with the radial profiles of the model and data {in the four cardinal directions. For the North and South profiles we sum along the shape of the bicone, while for the East and West profiles we sum excluding the bicone.} Our simple toy model is able to capture the general shape of the H$\alpha$ flux, along with matching the radial profiles quite well out to $\sim$2.5\,kiloparsecs. Our toy model is able to capture the shape of the kiloparsec scale feature visible in the southern part of the FP image, which comes from the intersection of the biconical structure with the disk of the galaxy at different position and inclination angles. Given the relative simplicity of our model and its ability to reproduce most of the AGN flux distribution in the data, we conclude it may be a viable solution for the kpc-scale AGN ionization seen in NGC\,1068.

\section{Summary \& Discussion}
\label{sec:conclude}

Using SALT RSS Fabry-P\'erot spectroscopy, we are able to map the [\ion{N}{2}] and H$\alpha$ complex across $\sim$2.6\,arcmin$^{2}$ of NGC\,1068. We measure the ionization state of the gas and, combined with SALT RSS longslit spectroscopy, {find that the kiloparsec-scale ionization features are powered by AGN photoionization, in complete agreement with \citep{dagostinoStarburstAGNMixingTYPHOON2018}.}
Our observations confirm the efficacy of using FP spectroscopy to study the extended effect of the AGN in nearby Seyfert II galaxies.

{We offer an alternative explanation for the source of the ionization features as photoionization due to past AGN activity. Our analysis suggest that the extended ionization seen in NGC\,1068 are due to enhanced AGN activity in the past and therefore the AGN history can be understood through spatially resolved studies of the ionized gas. Our toy model, while a relatively good fit to the data, is not definitive evidence that the ionization across the FoV directly traces the AGN's history. However, we believe our model presents a viable possibility which can be explored further in resolved future studies with detailed models that can appropriately back out the AGN intensity as a function of time from the distribution of AGN-ionized nebular gas.}

{We note that the variation in ionization and H$\alpha$ intensity across the FoV may be induced by other physical effects. Nebular gas density variations across the disk could potentially replicated the observed pattern. In addition, while optical spectroscopic analysis has limited what fraction of the ionization is induced by shocks in the kiloparsec-scale ionized features, this can be more directly measured from coronal emission line strengths (e.g. [\ion{Fe}{2}]$\lambda$1.257$\mu$m, [\ion{P}{2}]$\lambda$1.188$\mu$m, etc.), such as in \citet{teraoNearinfraredSpectroscopyNearby2016}, to place stricter limits on the contribution of shocks.}

$ $\\ % End line break due to AASTeX 6.3.1 errors with long acknowledgements when not using line numbers

We would like to thank the anonymous reviewer for their constructive comments which improved the final paper.

REH acknowledges support from the National Science Foundation Graduate Research Fellowship Program under Grant No. DGE-1746060. RCH acknowledges support from the National Science Foundation CAREER Award number 1554584. PV and RR acknowledge support from the National Research Foundation of South Africa.

{All of the observations reported in this paper were obtained with the Southern African Large Telescope (SALT) under proposal IDs 2011-2-RSA\_OTH-002 and 2015-2-SCI-02.}
This research made use of \texttt{saltfppipe}, a data reduction package for the SALT Fabry-P\'erot.

We respectfully acknowledge the University of Arizona is on the land and territories of Indigenous peoples. Today, Arizona is home to 22 federally recognized tribes, with Tucson being home to the O’odham and the Yaqui. Committed to diversity and inclusion, the University strives to build sustainable relationships with sovereign Native Nations and Indigenous communities through education offerings, partnerships, and community service.  

% \vspace{5mm}
\facilities{SALT (RSS)}

\software{Astropy \citep{collaborationAstropyCommunityPython2013}, Matplotlib \citep{hunterMatplotlib2DGraphics2007}, MPFIT \citep{markwardtNonlinearLeastsquaresFitting2009}, NumPy \citep{oliphantGuideNumPy2006,vanderwaltNumPyArrayStructure2011,harrisArrayProgrammingNumPy2020}, PyRAF/IRAF \citep{todyIRAFNineties1993,sciencesoftwarebranchatstsciPyRAFPythonAlternative2012}, PySALT \citep{crawfordPySALTSALTScience2010}, saltfppipe, SciPy \citep{virtanenSciPyFundamentalAlgorithms2020}, VorBin \citep{cappellariAdaptiveSpatialBinning2003}}

\bibliography{bibliography}{}
\bibliographystyle{aasjournal}
 
\end{document}